\documentclass{aa}

\usepackage{graphicx}
\usepackage[table]{xcolor}
\usepackage[dvipsnames]{xcolor}
\usepackage{txfonts}
\usepackage{multirow}
\usepackage{multicol}
\usepackage{tabu}
\usepackage{float}
\usepackage{natbib}
\usepackage[pdfauthor={},pdftitle={},colorlinks=true]{hyperref}

\definecolor{plt1}{HTML}{1f77b4}
\definecolor{plt2}{HTML}{ff7f0e}
\definecolor{plt3}{HTML}{2ca02c}
\definecolor{plt4}{HTML}{d62728}

\newcommand{\heim}{\textsc{heimdall}}
\newcommand{\dm}{\,pc\,cm$^{-3}$}
\newcommand{\psrcat}{\textsc{psrcat}}

\begin{document} 

   \title{The Northern High Time Resolution Universe pulsar survey}

   \subtitle{II. Single-pulse search set-up and simulations}

   \author{L.~J.~M. Houben\inst{1}\fnmsep\inst{2}\fnmsep\thanks{\email{l.houben@astro.ru.nl}}
           \and H. Falcke\inst{1}\fnmsep\inst{3}\fnmsep\inst{2}
           \and L.~G. Spitler\inst{2}
           \and \\ E.~D. Barr\inst{2}
           \and M. Berezina\inst{2}
           \and D.~J. Champion\inst{2}
           \and R. Karuppusamy\inst{2}
           \and M. Kramer\inst{2}\fnmsep\inst{4}}

   \institute{Department of Astrophysics/IMAPP, Radboud University, PO Box 9010,
              6500\,GL Nijmegen, The Netherlands
        \and
              Max-Planck-Institut f\"ur Radioastronomie, Auf dem H\"ugel 69,
              D-53121 Bonn, Germany
        \and
              ASTRON, the Netherlands Institute for Radio Astronomy, Oude Hoogevensedijk 4,
              7991 PD Dwingeloo, The Netherlands
        \and
              Jodrell Bank Centre for Astrophysics, University of Manchester,
              Alan Turing Building, Oxford Road, Manchester M13 9PL, UK}

   \date{}

   \abstract{The High Time Resolution Universe (HTRU) survey is an all-sky survey looking for pulsars and other radio transients. We present a new single-pulse (SP) search pipeline tailored to the northern part of the HTRU survey collected with the 100\,m Effelsberg Radio Telescope. In a selection of the data, synthetic SPs are injected with frequency-time structures resembling those of the detected fast radio burst population and processed by the pipeline to characterise its performance. Therefore, several new software toolkits have been developed ({\tt FRBfaker} and {\tt RFIbye}) to enable the injection of SPs with complex frequency-time structures and cope with the radio frequency interference (RFI) in the survey's data. We describe the operation of these toolkits alongside the overall functionality of the SP pipeline. Qualification of the pipeline confirmed that it is ready to process all the HTRU-North data. Additionally, we determine the survey's sensitivity to SPs, the impact of RFI thereon, the performance of the deep-learning classifier \textsc{fetch}, and some insights that may be used to improve the pipeline's performance in the future. Within the small data sample analysed, we detected 21 known pulsars and a rotating radio transient. In addition, eight faint SP trains that might originate from yet undiscovered neutron stars and 141 isolated SP candidates were discovered.}


   \maketitle

\section{Introduction}
The first discovery of pulsars \citep{hbp+1968} was made possible through the careful consideration of chart recordings that captured the individual single pulses (SPs) of these periodically radio-emitting neutron stars. Subsequently, Fourier searching techniques were developed that exploit their stable rotation and allowed for the detection of much fainter pulsars. However, SP searches continue to remain relevant and have led to the discovery of several new phenomena in pulsar astronomy. For example, a new class of rotating radio transients (RRATs) \citep{mll+2006} was found, which are a type of radio pulsar from which only infrequent emission is observed. Around the same time, a bright short-duration (several milliseconds) radio pulse of extra-galactic origin \citep{lbm+2007} was also found. This pulse turned out to be a new class of impulsive radio emission later referred to as fast radio bursts (FRBs). The search for SPs in radio imaging data even resulted in the discovery of yet another new phenomenon, that of long period transients (LPTs) \citep{hzb+2022,hrm+2023,clk+2024,hmb+2024,lcm+2025}. 

In particular, the detection of the first FRB caused a strong increase in interest in the search for transients in time-domain radio data. Archival data from several large radio telescopes were analysed for the possible presence of SPs with a high dispersion measure (DM), the integrated column density of free electrons along a particular line of sight. This resulted in the discovery of more of these elusive signals \citep{tsb+2013,cpk+2016,pab+2018} and the first repeating FRB \citep{ssh+2016}. With the potential of SP searching now clear, SP search capabilities were implemented on several telescopes, generally in real time. These efforts unlocked the discovery of even higher numbers of FRBs \citep{smb+2018,ffb+2019,taa+2021,vko+2023}, and with that, a new research field was born.

Finding new classes of short-duration transient phenomena, such as the ones just described, requires astronomical data with a high time and frequency resolution in order to not wash out their signals in noise. Surveys recording data in high resolutions are therefore ideal to search for these phenomena. One such survey is the High Time Resolution Universe (HTRU) survey, which is aimed to be an all-sky survey that searches for pulsars and other radio transients. It consists of a southern part (HTRU-South), with observations performed with the 64\,m Parkes Radio Telescope \citep{kjv+2010}, and a northern part (HTRU-North), with observations conducted with the 100\,m Effelsberg Radio Telescope \citep{bck+2013}, hereafter denoted with `Effelsberg'. The HTRU-South survey has been completed, and its data revealed a wealth of new sources, namely, the discovery of 244 pulsars \citep{mhth2005} (of which 42 are millisecond pulsars\footnote{\url{https://www.atnf.csiro.au/research/pulsar/psrcat/}} and 11 RRATs\footnote{\url{https://rratalog.github.io/rratalog/}}) and 28 FRBs \citep{tsb+2013,cpk+2016,pos+2018,tpp+2024}. The northern counterpart of the survey is still being conducted, and $\sim$16\% of the planned sky coverage has currently been observed. Quick analysis of roughly half of that data led to the discovery of 19 new pulsars, three of which are millisecond pulsars \citep{ber2019}.

With the creation of new and more sensitive radio telescopes in the northern hemisphere, including the Five-hundred-meter Aperture Spherical Telescope \citep{nlj+2011}, the likelihood that the remaining HTRU-North data contain yet unknown pulsars is reduced. The likelihood of detecting new short-duration (non-repeating) transients does not change, however. Moreover, other observatories might miss potential pulsars with blind searches, if they observe these pulsars at unfavourable times due to radio frequency interference (RFI) or the pulsar being in a nulling phase at the time of observation \citep{bac1970}. Re-observing these sky locations at different times and with different instruments could still reveal new sources. Hence, this paper reports on the creation of an SP search pipeline dedicated to searching the archival and upcoming HTRU-North data for short-duration transients. In contrast to the HTRU-South survey, no detailed SP search has been performed on the HTRU-North data, leaving an opportunity to detect transients and possible new transient phenomena in these data. To determine the pipeline's capabilities of characterising the time variable radio sky down to sub-millisecond time resolution, we performed SP injection tests with SPs of varying morphologies (the flux distribution over frequency and time) resembling those found for FRBs \citep{pgk+2021}.

In the following sections, the selection of survey data used to characterise the new SP pipeline is first described. Thereafter, we  give an overview of the pipeline. In Sect. \ref{sec:analysis} we elaborate on how synthetic SP injections are used to determine the pipeline's performance. The results obtained from the analysis of the first HTRU-North data sample are presented in Sect. \ref{sec:results}, and then future prospects for the pipeline are presented in Sect. \ref{sec:discussion}. These prospects can be used to aid forthcoming SP searches and the further analysis of the HTRU-North data. Sect. \ref{sec:conclusions} concludes the paper, outlining all the major findings.

\section{Observations and data reduction} \label{sec:obs&datareduction}
\subsection{HTRU-North survey}
The HTRU-North survey's observations are performed with Effelsberg's 21\,cm multibeam receiver. This receiver consists of seven horns, each capable of recording  300\,MHz of bandwidth (BW) centred on 1360\,MHz in two polarisation channels. For the HTRU-North survey, these channels are always summed to attain total intensity data. This is done after the Effelsberg Pulsar Fast Fourier Transform Spectrometer backend digitised the analogue signal of each of the receiver's horns and processed them with a 512-channel polyphase filterbank. For each beam, a \textsc{sigproc}\footnote{\url{https://sigproc.sourceforge.net/}} filterbank file with 512 frequency channels and a time resolution of 54.61\,\textmu s is obtained. These are then downsampled to an 8-bit precision with the dedicated code {\tt 421}.

The seven beams allow for an efficient coverage of the sky by pointing consecutive observations such that the gaps of the previous pointing's (PT) beam pattern are covered by the next (see Fig. 2 in \citet{bck+2013} for a visual explanation of the tiling performed). With this tiling, the northern hemisphere, above a declination of \mbox{-20$^{\circ}$}, can be fully observed with about 218\,000 PTs. The declination limit here arises from the physical location of Effelsberg. The HTRU-North's PTs are also grouped into three ranges based on the Galactic latitude of the centre beam of the PTs. These ranges are observed with integration times that are chosen such that for each latitude range, the HTRU-North survey has similar sensitivity limits as the HTRU-South survey.

The PTs covering the sky with Galactic latitudes of |$b$| > 15$^{\circ}$ are observed with integration times of 90\,s, the survey's high-latitude (high-lat) range. The mid-latitude (mid-lat) range covers Galactic latitudes of |$b$| < 15$^{\circ}$ with 180\,s integration times. The PTs covering the range of |$b$| < 3.5$^{\circ}$ with long integrations of 1500\,s constitute the low-latitude (low-lat) section of the HTRU-North survey. With these long low-lat integration times, the survey aims to probe deep into the Galactic plane to discover many faint pulsars. 0.46\% of the HTRU-North's low-lat PTs, 53\% of its mid-lat PTs and 4.8\% of the high-lat PTs have currently been observed.

\subsection{Data sample}
The low-lat region is also covered with shorter mid-lat PTs, as can be seen from the latitude ranges above. A shallow sweep of the Galactic plane is so performed with which bright pulsars can quickly be discovered without having to wait for the long integration observations of the actual low-lat PTs. The mid-lat PTs thus consist of a homogeneous dataset of observations in and outside the Galactic plane. Together with a 53\% completion rate, they present a perfect dataset to characterise the pipeline. For that purpose, a total of 1500 mid-lat PTs were selected across the Galactic sky (see Fig. \ref{fig:pointing-skymap} for their Galactic distribution). With the first 500 PTs, a pre-analysis was performed to determine appropriate parameters for the software tools used in the SP pipeline.

The other 1000 PTs were used for the injection analysis described in Sect. \ref{sec:analysis} and are referred to as the analysis PTs. Half of these PTs were chosen with a latitude within the low-lat range (i.e. |$b$| < 3.5$^{\circ}$) and chosen so as to be approximately uniformly distributed over the celestial sky (Fig. \ref{fig:rfi-skymap}). These PTs shall be referred to as the `quick low-lat PTs'. The other 500 mid-lat PTs were selected with a Galactic latitude of |$b$| > 3.5$^{\circ}$ and |$b$| < 15$^{\circ}$, and thus exclude the Galactic plane. Whenever in the following `mid-lat PTs' is mentioned, this refers to these PTs with a |$b$| > 3.5$^{\circ}$. They are chosen similarly to the quick low-lat PTs to be evenly distributed over the celestial sky too. The selected data samples thus cover many different telescope orientations to resemble those of the full survey and probe the entire RFI environment around Effelsberg in which the HTRU-North observations are conducted.

\subsection{Processing pipeline}\label{sec:pipeline}
A dedicated SP pipeline was developed to search the HTRU-North data for transients and possible new transient phenomena. The pipeline is designed to run on a supercomputer equipped with graphics processing units (GPUs). Here, the pipeline ran on the compute cluster \textsc{hercules}\footnote{\url{https://docs.mpcdf.mpg.de/doc/computing/clusters/systems/Radioastronomy.html}} of the Max Planck Computing and Data Facility (MPCDF) to have easy access to the archived HTRU-North data at the MPCDF. The workflow of the pipeline is outlined in Fig. \ref{fig:sp-pipeline} and consists of the following consecutive operations.

First, a selection of PTs is recalled from the MPCDF's tape archive and copied to \textsc{hercules}. There, each PT's tarball is unpacked to allow its seven beam files to be inspected for copy completeness and searched for SPs.
    
Second, the SP search is performed with \heim\footnote{\url{https://sourceforge.net/projects/heimdall-astro/}}\textsuperscript{,}\footnote{A version of \heim\ is used that includes the updated root mean square (rms) calculation, as suggested by \citet{gff+2021}, to make \heim\ return more accurate S/N values.} \citep{bbbf2012} to make it complementary to that of the HTRU-South. Most of the FRBs in that survey were discovered with this GPU-based software package. As has been pointed out by \citet{qkb+2023}, \heim's DM tolerance parameter has a significant effect on its capability to detect dispersed signals. The {\tt --dm\_tol} flag determines the DM step size between each DM trial within the set DM search range. Setting it to a high value means that \heim\ performs a coarse search over DM and so potentially misses SPs with a DM in between two DM trials. Therefore, this flag is set to a lower tolerance of 1.05 compared to the default value of 1.25.

Similar to \citet{cpk+2016}, the data are also searched over a pulse width range of 2 to 256 samples to reduce the number of candidates produced by \heim, which quickly increase in number with an increased search width. Since pulses with low DMs are likely to originate from terrestrial sources, SPs are searched with DMs greater than 10\dm. \heim\ is further set to search for candidates with an S/N $\geq$ 6.5. With this threshold and the other set search parameters, less than one false positive detection is expected in a beam file of a PT, due to Gaussian noise (see Sect. \ref{sec:RFI-env}). Together with Effelsberg's size, and thus sensitivity, the applied S/N threshold allows for the detection of very faint signals. In this manner, each beam file is searched for SPs over a range of pulse widths of 0.109 to 13.98\,ms and over 2200 DM trials ranging from 10 to 5000\dm.
    
Third, RFI is mitigated. Effelsberg is located near densely populated areas and can be heavily affected by RFI. To remove this RFI, a custom RFI mitigation code was made ({\tt RFIbye.py}) that preserves more of the data than existing RFI mitigation tools might do. In this way, the number of candidates found by \heim\ due to RFI is reduced, and the detection of fainter signals is enabled. The code identifies affected data samples using several thresholding techniques and replaces them with noise indistinguishable from the non-affected data. The sample replacement is then applied to a copy of the beam files. From each channel of these files, the channel's median is subtracted and divided by the channel's standard deviation (STD) to flatten the files' bandpass. Sect. \ref{sec:RFIbye} describes in more detail how {\tt RFIbye.py} is used.

Fourth, once the cleaned filterbank beam files are obtained, \heim\ is used to search them again for SPs. This is done similarly as described in the second step.

Next, the SP candidates are further reduced in number by selecting those candidates that pass the following criteria:
\begin{align*}
    N_{m} &\geq 3,\\
    W &> W_{DM_{smear}},
\end{align*}
where $N_{m}$ is the number of boxcars or DM trials that were grouped into a single candidate by \heim\ and $W_{DM_{smear}}$ is the pulse width a candidate would minimally have due to dispersion smearing in the data's highest frequency channel. The criteria are imposed with an adaptation of \heim's {\tt trans\_gen\_overview.py} that performs its candidate classification. In the following, candidates that pass these criteria are referred to as `valid candidates'.

Subsequently, to ease the manual assessment of the resulting candidates, a custom plotting tool was developed, inspired by {\tt trans\_gen\_overview.py}. This tool, further described in Appendix \ref{sec:cand_viewer}, presents an overview of the candidates per beam and the union of candidates of all beams. For the latter, candidates detected in two or more non-adjacent beams are classified as RFI when they are not spaced further apart in time than three times their maximum width. This classification should, however, be taken as an indication of the candidate's origin since the time is not synchronised across the beams of Effelsberg's 21\,cm multibeam receiver.

To obtain another indication of whether a candidate is a true positive, they are passed onto \textsc{fetch} \citep{aab+2020} to obtain a score resembling the probability with which this deep-learning classifier determined a candidate to be a true transient. \textsc{fetch} provides eleven network architectures (models $a$ to $k$) to calculate these probabilities with. Each candidate is evaluated by every model. If the averaged probability of all eleven models exceeds 0.5, the candidate is logged as being detected by \textsc{fetch} in a separate text file together with the letters of the models that returned a probability of $\geq$0.5 for that candidate. The averaged probability of all models is taken to make \textsc{fetch}'s classification more robust, which has been recommended by \citet{aab+2020}. In this way, the human assessments are made more reliable whenever the \textsc{fetch} results are consulted during the human evaluation of candidates, because the visual inspection of a candidate proved indecisive. These results then provide another indication of the authenticity of a candidate.

Finally, the candidates deemed real through human inspection are added to the SQLite database of the HTRU-North SP search that keeps track of the progress and results of the search. For candidates with an S/N > 8, their PT's beam files and \heim\ candidate lists are saved.

In contrast to other surveys \citep{als+2020,chg+2022,myw+2022,tpp+2024}, \textsc{fetch} is not used to pre-select candidates for human evaluation in the second to last step, but as a tool to guide this evaluation. This is done to allow for the discovery of unexpected signals, which would have potentially been selected against by \textsc{fetch} as its models are trained to detect FRB-like transients. New transient phenomena need not be FRB-like.

\subsection{RFIbye} \label{sec:RFIbye}
{\tt RFIbye.py} is a Python script designed to remove RFI from filterbank data by replacing all data samples that exceed a given set of thresholds\footnote{For details on the possible thresholds consult the extended \textsc{readme} in the GIT repository of {\tt RFIbye.py}: \url{https://gitlab.com/houben.ljm/rfibye}} with randomly selected data samples unaffected by RFI on a channel-by-channel basis. The cleaned data are then saved as a copy of the original filterbank file. In this manner, every software toolkit capable of reading filterbank files can be used to analyse the cleaned data.

To determine which of {\tt RFIbye.py}'s thresholds to use and at what value they need to be set for the HTRU-North data, the 500 pre-analysis PTs were cleaned with different sets of thresholds. {\tt inspectra.py}, part of the {\tt RFIbye} toolkit, was then used to find the most balanced set of thresholds that removed the manually identified RFI from the data while replacing the least amount of data samples.

During this investigation, it was noted that some beam files were affected by a peculiar type of RFI that is visible as an oscillation in the data’s frequency-averaged time series (TS) with a fixed period. Even though these periodic signals are seen in the undispersed data, they are reconstructed when the DM sweep for a de-dispersion step is a multiple of the oscillation's period. This causes \heim\ to report many candidates with a wide variety of DMs. Hence, a routine was added to {\tt RFIbye.py} to identify this type of RFI specifically and remove it from the data.

For the HTRU-North SP pipeline, {\tt RFIbye.py} mitigates RFI in blocks of data of 2$^{15}$ spectra at a time. For each block, the six highest and lowest frequency channels are masked to account for possible band roll-off ({\tt --rm\_band\_edge}). Then the median and STD are calculated for the data block’s bandpass, the full resolution TS, the TS downsampled by a factor of eight, and the spectrum for each time bin. ({\tt RFIbye.py} arguments {\tt --zap\_chan\_thres} 6, {\tt --ts\_thres} 5, {\tt --ts\_downsampfrac} 3 and {\tt --std\_thres} 6, respectively). From these distributions, bad spectra and channels are identified if their power deviates more from the distribution's median than the set threshold times the distribution's STD. If such spectra or channels are found, they are masked for later replacement with random unmasked data samples. The TS of the data block is also autocorrelated to find a possible period of periodic RFI within the block. If such periodic RFI is found, all spectra over the time span of the RFI are masked ({\tt --rm\_reg} 0). Lastly, five additional spectra around each masked spectrum are masked, in order to remove spectra that may be affected by an RFI bleed from neighbouring bad spectra, but do not themselves exceed the set thresholds ({\tt --edge} 5).

A mask is obtained that indicates which spectra and channels are considered bad through the application of the user-defined thresholds. If this mask contains spectra for which more than 30\% of its channels are masked, the entire spectrum is masked too. Likewise, a channel is fully masked when more than 20\% of its time-samples are masked. These percentages were determined during the analysis of the pre-analysis PTs, resulting in the detection of the fewest false positive events by \heim. Per channel, masked elements are replaced by randomly drawn unmasked elements to retain the noise distribution of the channels. In the case that a channel is fully masked, it is replaced by Gaussian noise with a mean and STD determined from the smoothed bandpass filling the missing entries of the fully masked channels.

After replacing all masked samples with valid samples, the mean of each spectrum is subtracted from the data block to apply a zero-DM filter as described by \citet{ekl2009} ({\tt --zerodm}). The data block's bandpass is determined through fitting a line to the time-averaged channels and then subtracted from the data ({\tt --rm\_bandpass}). With this subtraction, the resulting channels are divided by each channel's STD measured over the block of data processed ({\tt --chan\_indep}). The cleaned data block is stored in a copy of the original filterbank file, and the next is read to iteratively remove RFI from the entire data file.

The actual removal of RFI from filterbank files by {\tt RFIbye.py} enables users to more efficiently deal with RFI during the analysis of their data. For instance, instead of providing software tools with lists of bad frequency channels, they can just be removed with {\tt RFIbye.py} (using {\tt --zap\_chans}), so they need not be accounted for in every analysis step. This might prove especially useful in cases where the routine {\tt rfifind} of the popular toolkit \textsc{presto}\footnote{\url{https://github.com/scottransom/presto}} \citep{ran2001} has been used to identify RFI, and it is desirable to perform further analysis with another toolkit. With {\tt RFIbye.py}, a created {\tt rfifind} mask can be applied to a filterbank file (using {\tt --rfifind\_mask}) in order to continue its analysis outside of \textsc{presto} with a cleaner version of the file. In the case of the HTRU-North SP pipeline, the cleaned data files can be searched again with \heim\ for SPs.

\subsection{Search sensitivity}\label{sec:search_sensitivity}
\begin{figure}
\centering
\includegraphics[width=0.48\textwidth]{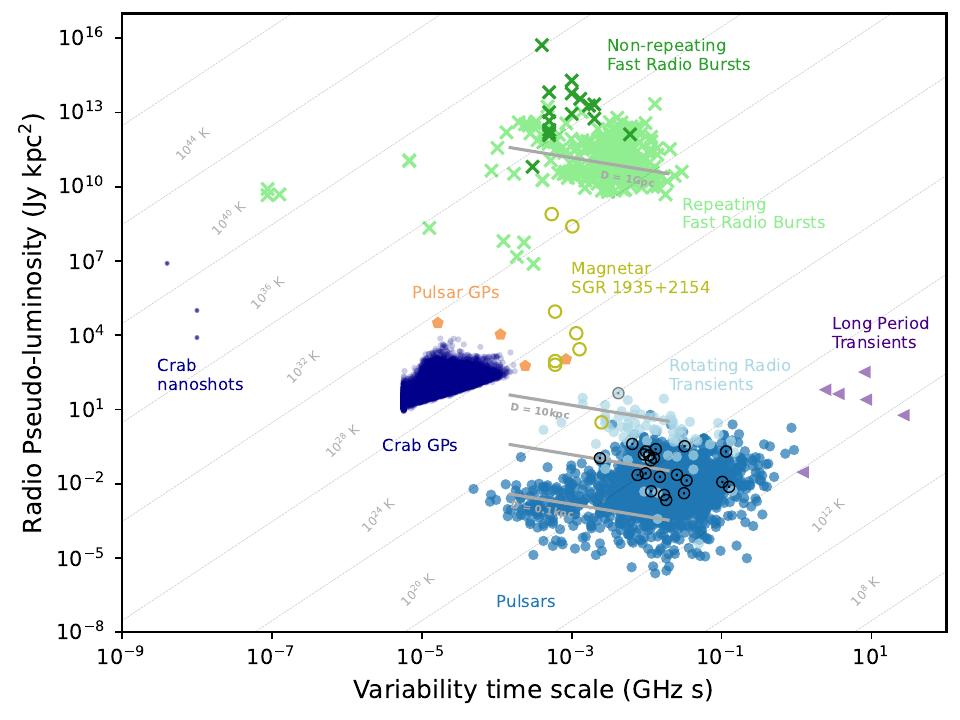}
   \caption{Transient phase space showing the radio luminosity versus the variability timescale of several transient events observable in the radio sky. Overlaid are Effelsberg's sensitivity curves for the Galactic distances of 0.1, 1, and 10\,kpc as well as the cosmological distance of 1\,Gpc and pulse widths from 0.109 to 13.98\,ms (the used search width range of the SP search). The positions of the re-detected pulsars and RRAT are indicated with the black and grey $\odot$ symbols, respectively.}
   \label{fig:tps}
\end{figure}

The HTRU-North data are blindly searched for SPs. To guide the manual inspection of the resultant candidates, it is helpful to have an idea of what kind of sources are expected to be found. Effelsberg's theoretical sensitivity limit is therefore calculated for the detection of an SP with a width, $W$, of 1\,ms using the radiometer equation \citep{mc2003}
\begin{equation}\label{eq:radiometer}
    S_{max} = \frac{S/N \cdot S_{sys}}{\sqrt{n_{p}\Delta\nu W}}.
\end{equation}
For the HTRU-North data, the number of summed polarisations, $n_{p}$, effective bandwidth, $\Delta\nu$, and system equivalent flux density (SEFD), $S_{sys}$, are $n_{p}$ = 2, $\Delta\nu \approx$ 290\,MHz, and $S_{sys} \simeq$ 15\,Jy. The SEFD is taken from the 21\,cm multibeam receiver documentation. Using the S/N threshold of 6.5 from the SP pipeline, an SP fluence limit is found of \mbox{$F_{lim} \simeq$ 0.13\,Jy\,ms}.

This fluence limit can be used to draw sensitivity curves for specific distances on top of the transient phase space (TPS). In Fig. \ref{fig:tps}, this is done for Galactic distances of 0.1, 1, and 10\,kpc, as well as the cosmological distance of 1\,Gpc, and pulse widths from 0.109 to 13.98\,ms. From this figure, it follows that the performed SP search is, in theory, sensitive to a large range of transient events. We may expect to find SPs of pulsars and RRATs (Sect. \ref{sec:re-detections}) and possibly giant pulses of pulsars, FRB-like pulses of magnetars, or even several FRBs. These expectations form promising prospects for the full analysis of the HTRU-North data, as the chance of detecting sporadic transient events greatly increases with the total amount of data processed.

\section{Analysis} \label{sec:analysis}
\subsection{Single-pulse injections}
Expectations about what SPs are expected to be found can be further refined by determining the HTRU-North SP pipeline's efficiency and completeness for the detection of SPs from a specific parameter space. For the newly implemented SP search pipelines at other radio observatories, this is commonly done with SP injection tests \cite[e.g.][]{als+2020,gff+2021,myw+2022,mts+2023,qkb+2023}. While most injections have been limited to simple Gaussian envelopes that show little structure in frequency, \citet{pgk+2021} showed that at least the Canadian Hydrogen Intensity Mapping Experiment (CHIME) repeating FRBs exhibit quite a complex frequency structure. They identified four archetypical morphologies (I to IV) of FRB bursts based on the FRBs published in the first CHIME/FRB catalogue \citep{taa+2021}.

Bursts of morphology I (MI) represent single Gaussian envelopes spanning the full BW. Those of morphology II are single Gaussian envelopes with constrained BWs. A burst represents morphology III if it is composed of multiple sub-bursts of morphology I or II that are closely spaced apart in time. Similarly, a burst belongs to morphology IV if its sub-bursts are of morphology II and they are both spaced apart in time and frequency. Fig. \ref{fig:A.eff}, \ref{fig:A.chime} and \ref{fig:A.arecibo} are examples of how bursts of morphology type II (MII), III (MIII) and IV (MIV) look like. Together with bursts from MII, MI bursts closely resemble pulses from other known SP-emitting sources. Identifying the completeness of the pipeline for the archetypical FRB morphologies thus gives a good impression of its capability to detect a range of sources.

\subsubsection{Injection parameter space}\label{sec:inj_param_space}
For the characterisation of the pipeline's completeness, synthetic SPs were created with a wide range of burst parameters and injected into the beam files of the selected analysis PTs. This large variation in burst parameters allows for the investigation of potential conditions under which the pipeline unexpectedly detects or misses injections. The parameters are chosen such that the applied injection analysis mimics a real blind search for SPs. The ranges of burst parameters therefore ensure that part of the injections are hidden within the noise of the data and only their brightest constituents are detectable through the utilisation of a blind SP search. The parameters define the injections' intrinsic properties, i.e. they specify the injections' morphology before propagation effects, such as scattering and dispersion, are applied, which might deform their morphology.

In each of the HTRU-North analysis PTs, four synthetic SPs are injected in four randomly chosen beam files of a PT. Every injection of a PT is assigned a specific set of burst parameters to give it a frequency-time structure resembling that of one of the FRB morphologies. Each PT thus contains an injection of each morphology. In Appendix \ref{sec:A.inj_params}, the parameter assignment is discussed in detail and the extent of the assigned parameters is visually depicted in Fig. \ref{fig:inj_params}. To assure a large morphological variation among the injections of an FRB morphology in the data, their burst parameters are randomly drawn from distributions summarised in Table \ref{tab:injectionparams}. The choice of these distributions is therefore data driven, and they do not reflect intrinsic physical properties of the FRB population. The sets of burst parameters are stored in an SQLite database together with the randomly picked magnitudes of the propagation effects that should be applied to the injections.

The custom code, {\tt FRBfaker}, has been written to construct synthetic SPs from these stored parameters and inject them at arbitrary times within the beam files of the HTRU-North analysis PTs. Therefore, this code is temporarily incorporated in an additional step inserted between steps 1 and 2 of the SP pipeline (Sect. \ref{sec:pipeline}).

\begin{table}
\caption{Overview of distributions, per FRB morphology, from which the observational parameters are drawn for the injection test.}
\label{tab:injectionparams}
\centering
\begin{tabular}{l c c c c l}
    \hline
    \noalign{\smallskip}
    Morphology & I & II & III & IV & \\
    \noalign{\smallskip}
    \hline
    \noalign{\smallskip}
    Fluence    & \multicolumn{4}{c}{\cellcolor[gray]{0.9} 0.02 - 6.1\,Jy\,ms\;$^{a}$} & \\
    DM   & \multicolumn{4}{c}{\cellcolor[gray]{0.9} 1.5 - 5,000\dm\;$^{b}$} & \\
    $\tau_{scatt}$  & \multicolumn{4}{c}{\cellcolor[gray]{0.9} 8.56e$^{-7}$ - 10.11\,s\;$^{c}$} & \\
    subC\_width & \multicolumn{4}{c}{\cellcolor[gray]{0.9} 54.61\,\textmu s - 10\,ms\;$^{d}$} & \\
    \noalign{\smallskip}
    $N_{comp}$ & 1 & 1 & >\,1\;$^{e}$ & >\,1\;$^{e}$ & \\
    BW         & 300\,MHz & \multicolumn{3}{c}{\cellcolor[gray]{0.9} 10 - 300\,MHz\;$^{f}$} & \\
    $f_{ref}$  & 1360\,MHz & \multicolumn{3}{c}{\cellcolor[gray]{0.9} 1210 - 1510\,MHz\;$^{g}$} & \\
    subC\_time  & - & - & equal\;$^{h}$ & equal\;$^{h}$ & \\
    drift-rate & - & - & 0 & $^{i}$ & \\
    \noalign{\smallskip}
    \hline
\end{tabular}
\tablefoot{$a$) Power law distr. with $\alpha$ = -1.5 \citep{msb+2019}, where the fluence limits originate from the radiometer equation for a burst with a width of 54.61\,\textmu s (Effelsberg's sampling time) and an S/N of 4 ($F_{min}$) and a width of 100\,ms and an S/N of 100 ($F_{max}$). ~$b$) Exponentially modified Gaussian distr. with its parameters matching those of the DM distribution of the known FRB population. ~$c$) Log-uniform distr. referenced to 1\,GHz and set to 0 whenever $\tau_{scatt}$ is smaller than 20\% of the dispersion smeared burst width. ~$d$) Burst subcomponent widths are drawn from a log-uniform distr. ~$e$) Log-uniform distr. from 2 to 12 components. ~$f$) Log-uniform distr. ~$g$) Uniform distr. ~$h$) The time between bursts' subcomponents are drawn from a uniform distr. from subC\_width to subC\_width $\cdot$ $N_{comp}$. ~$i$) Uniform distr. from BW$_{chan}$/subC\_time to full BW/(subC\_time $\cdot\ N_{comp}$), where $f_{ref}$ is taken into account to ensure that subcomponents do not fully drift outside the band.}
\end{table}

\subsubsection{FRBfaker}\label{sec:FRBfaker}
{\tt FRBfaker} was created to allow for the injection of a single SP with a user-defined morphological complexity into filterbank-type data. Users can construct synthetic copies of any observed SP and inject it either in a provided filterbank file or in a fake filterbank file made up of pure Gaussian noise. The code is publicly available\footnote{\url{https://gitlab.com/houben.ljm/frb-faker}} and extensively tested as described in Appendix \ref{sec:A.faker-tests}. For its inclusion in the HTRU-North SP pipeline, it has been written such that it accepts, among others, the observational parameters from Sect. \ref{sec:inj_param_space} to define the morphology of an injection. Moreover, a custom dynamic spectrum (frequency-time profile) can also be provided to the {\tt FRBfaker} if the accepted observational parameters do not provide enough flexibility to inject a desired morphology. It then takes care of the appropriate scaling and propagation effects before injecting the synthetic SP.

The scaling of a pulse is necessary to give it the requested fluence or S/N in its frequency-averaged TS. Whenever an injection must be performed in integer data, such as in the case of the HTRU-North data, special attention must be given to this scaling. The high fidelity with which a synthetic burst is created might then be lost due to the limited bit-range of the data (see Sect. \ref{sec:D-discrepancies} and Appendix \ref{sec:A.int_inject}).

For the injection analysis presented here, bursts were injected with a given intrinsic fluence, i.e. their fluence prior to propagation effects taking effect. To connect these fluences with data properties, the {\tt FRBfaker} turns the SP fluences into S/N values with Eq. \ref{eq:radiometer} and makes use of the SEFD, $n_{p}$, and BW of Effelsberg's 21\,cm multibeam receiver. Then, Eq. \ref{eq:radiometer} is used again to calculate what area the injections' pulse profiles should have in order to give them the desired fluence in the data. Therefore, the profiles are assumed to be Gaussian shaped, and the rms of the data's noise is used as $S_{sys}$. The injections' profiles are then scaled accordingly before they are injected into the data. In Appendix \ref{sec:A.faker-tests}, this scaling method is referred to as the `original {\tt FRBfaker} S/N scaling'.

\subsection{Injection recoveries}\label{sec:inj-recov}
The recovered injections can give insight into the efficiency of the pipeline to detect SPs. It is therefore most convenient to be able to directly compare the pulse parameters reported by \heim\ with those with which they were injected. The synthetic bursts were injected with a given intrinsic fluence, while \heim\ reports S/N values. Post-injection, all injected SPs have therefore been re-made with the {\tt FRBfaker} to calculate with which S/N they were injected after propagation effects were applied. This is done through convolving the remade bursts with boxcars of varying widths that were increased with one time-sample at a time. The boxcar width that resulted in the highest deduced S/N was taken to be the propagated width of the injected SPs and the corresponding S/N as its injected S/N. In other words, the `boxcar convolution' method (discussed in Appendix \ref{sec:A.scaling_test}) was applied to deduce the injections' injected S/Ns, since the so calculated S/Ns resemble \heim's recovered S/Ns most closely.

However, \heim\ searches for SPs with discrete steps of $2^{j}$ time samples, where $j$ ranges from one to eight. This can result in a mismatch between \heim's applied boxcar width and the actual width of the burst. \heim's recovered S/N values will therefore differ from the injected S/N values calculated above. To assess the effects of the mismatch between the applied search boxcar widths and the propagated pulse widths, for all bursts, the S/N values were also calculated using the best-fitting \heim\ boxcar width. These S/N values are referred to as an injection's `discrete injected S/N'.

For the injection analysis, one thousand synthetic SPs were injected per morphology. The wide ranges of assigned intrinsic burst parameters caused these bursts to have a large spread in injected S/N too. Since the S/N of Gaussian noise is one, injections with such an injected S/N or smaller are indistinguishable from noise and not expected to be detectable. Therefore, only injections with an S/N$_{inj}$ > 1 are taken into account. In this manner, a total of 2717 synthetic SPs were included in the analysis that have randomly chosen parameters drawn from the distributions listed in Table \ref{tab:injectionparams}. Table \ref{tab:num_recoveries} lists the number of included injections per morphology.

\begin{table}
\caption{Number of included and detected injections per morphology.}
\label{tab:num_recoveries}
\small
\centering
\begin{tabular}{l c c c c c}
    \hline
    \noalign{\smallskip}
    Morphology & I & II & III & IV & Total \\
    \noalign{\smallskip}
    \hline
    \noalign{\smallskip}
    $N_{S/N_{inj}\ >\ 1}$ & 825 & 697 & 576 & 619 & 2717 \\
    $\rightarrow N_{t_{inj}\ <\ 175\,s}$ & 805 & 677 & 569 & 603 & 2654 \\
    \noalign{\medskip}
    Detections & 236 & 216 & 155 & 188 & 795 \\
    \multicolumn{1}{r}{$^{S/N\ >\ 6.5}$} & 29.3\% & 31.9\% & 27.2\% & 31.2\% & 30.0\% \\
    \noalign{\smallskip}
    \hline
\end{tabular}
\end{table}

\subsubsection{Pipeline detections and visual inspections}\label{sec:detect&ident}
Of the injected synthetic SPs, 795 were blindly detected with the HTRU-North SP pipeline. Twenty-four injections with a discrete injected S/N above 6.5 (\heim's S/N detection threshold), and therefore expected to be detectable, were missed and visually inspected to determine a reason as to why they were not detected. The discrete injected S/N was used here since \heim\ searches with discrete boxcar widths that might not recover all of an injection's power. The discrete injected S/Ns should encompass potential mismatches between the search boxcar width and the actual width of an injection and can thus be considered the most ideal recoverable S/Ns by \heim\ in the absence of RFI.

Nine injections were identified that should have easily been detectable by the pipeline but were nevertheless missed. They were injected outside of the time range searched by \heim, which only searches for SPs in a time range that can fully contain the maximum DM delay over the data's frequency range. No information can therefore be obtained about the performance of the SP pipeline from bursts injected at a time $\gtrsim$175\,s in the analysed PTs. These are consequently ignored for the remainder of the analysis (see Table \ref{tab:num_recoveries}).

When the above injections are not taken into account, the remaining non-detections are due to the injections having been injected coincident with an RFI spike or due to chance (for the latter see Sect. \ref{sec:D-discrepancies}). The former happened once and shows the proper functionality of {\tt RFIbye}. The RFI spike on top of which the synthetic SP was injected was removed by {\tt RFIbye} and the SP itself remained in the data. However, its power was significantly reduced due to the removal of the RFI spike. This caused it to be only visually identifiable and not detected by \heim. All injections that were potentially detectable according to their discrete injected S/N were actually visually identifiable in the data, showing that {\tt RFIbye} has not removed any of these injections from the data.

\subsubsection{Completeness and recall}\label{sec:pipeline-recall}
\begin{figure}
\centering
\includegraphics[width=0.5\textwidth]{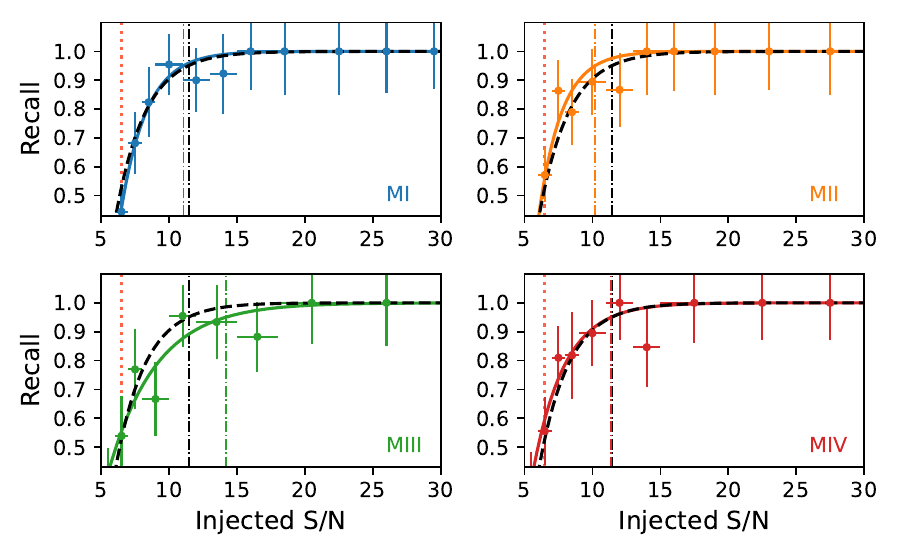}
   \caption{Recall curves for each of the four morphologies as a function of the SPs' injected S/Ns. The injected S/Ns for which the HTRU-North SP pipeline is complete are indicated by the coloured dash-dotted lines and entail $\sim$11 for MI, $\sim$10 for MII, $\sim$14 for MIII, and $\sim$11 for MIV. The morphology averaged recall is plotted with dashed black lines. The S/N above which the HTRU-North SP pipeline is considered to be complete, irrespective of a burst's morphology, is $\sim$11.}
   \label{fig:snr-recall}
\end{figure}

Comparing blindly detected injections with the total number of injections within a specific S/N range reflects the pipeline's ability to find SPs in that range. To characterise the efficiency of the HTRU-North SP pipeline, this metric (also known as recall) is calculated. The injections are therefore, per morphology, binned with widths of one S/N unit. If the bin contains fewer than ten injections, it is merged with the following S/N bin until a bin is formed with at least ten injections. Next, the recall of the S/N bin is calculated by dividing the number of blindly detected SPs in the bin by the total number of injections in that bin. The error on this value is taken to be $1/\sqrt{N_{bin}}$, where $N_{bin}$ is the total number of injections in the S/N bin. The recall values against the injected S/N of the centre of the bins is shown in Fig. \ref{fig:snr-recall}. Exponential curves of the form
\begin{equation}\label{eq:exponential}
    c - a\,e^{-b\,x}
\end{equation}
are fitted to the calculated recall values per morphology and displayed as the solid coloured curves. The dashed black line represents the exponential fit through the recall values of all the injections combined. From these fits, an injected S/N can be deduced for which the HTRU-North SP pipeline is complete. If the pipeline is considered to be complete for a recall value of 0.95, i.e. when it detects at least 95\% of the SPs in the data, the pipeline is complete for an S/N of $\sim$11, $\sim$10, $\sim$14, and $\sim$11 for morphologies I, II, III, and IV, respectively. A completeness limit of an S/N of $\sim$11 is found when considering all the injections together, irrespective of their morphology.

\begin{figure}
\centering
\includegraphics[width=0.5\textwidth]{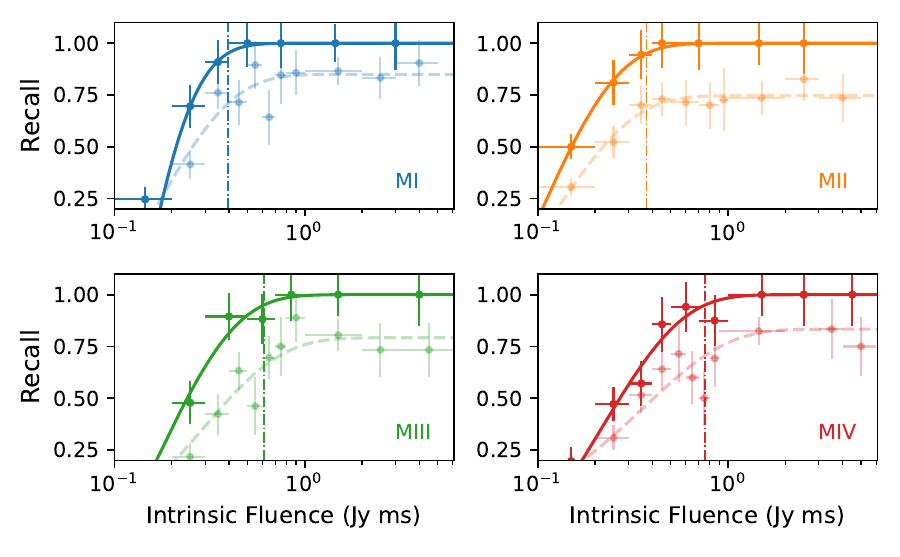}
   \caption{Recall curves as a function of the SPs' intrinsic fluences per morphology. When all injections are included, these recall curves (dashed pale lines) converge around a value of 0.8 ($c$ in Eq. \ref{eq:exponential}). This convergence occurs mainly due to the applied scattering, rendering injections undetectable even for the brightest injections. By excluding all scattered injections, the solid recall curves were obtained from which fluence completeness values for the pipeline follow that are $\sim$0.4, $\sim$0.6, and $\sim$0.8\,Jy\,ms for morphologies I/II, III, and IV, respectively.}
   \label{fig:fluence-recall}
\end{figure}

The intrinsic fluence of an SP is, however, the physically more interesting parameter to obtain. Thus, to determine from which intrinsic fluence the pipeline is complete, the above process is repeated, binning the injected SP fluences. This yields the result depicted in Fig. \ref{fig:fluence-recall} as the pale dashed lines. For the probed fluences, the SP pipeline never reaches a recall value of 1. The injected fluences are not high enough to always render the injections detectable, meaning that something reduces their recoverable S/N. Since the {\tt FRBfaker} scales injected SPs such that they have a given fluence or S/N in the data's TS, limitations in an SP's BW are accounted for by the {\tt FRBfaker}. The reduction in the recoverable S/N must therefore be caused by width mismatches between the applied search boxcar widths by \heim\ and the actual (propagated) widths of the bursts.

The discrete injected S/N of all injections are plotted against their injected S/N in Fig. \ref{fig:wth_dependence} to investigate what burst properties might induce such mismatches and potentially render an injection undetectable. In this figure, the markers, their colours and sizes represent the number of subcomponents an injection consists of, the potential scattering applied to it, and the width mismatch (the larger the size, the larger the mismatch), respectively. As can be seen from the deviations from the dashed grey line, representing the situation where \heim's boxcar widths perfectly probe the injections' burst widths, the largest mismatches are mainly caused by the imposed scattering times that widen the injections far beyond the maximum searched width of 13.98\,ms. The recoverable S/N of even the brightest injections can be reduced to such a degree that they become undetectable.

A value for the fluence completeness of the pipeline can therefore best be obtained through the consideration of the unscattered injections only. Hence, the solid curves in Fig. \ref{fig:fluence-recall} that represent the pipeline's recall per morphology deduced from the injections without any scattering applied to them. From these curves, a fluence can be read off for which the pipeline reaches a recall of 0.95 and is considered to be complete: $\sim$0.4\,Jy\,ms for MI and MII, $\sim$0.6\,Jy\,ms for MIII, and $\sim$0.8\,Jy\,ms for MIV.

The pipeline is complete for a higher intrinsic fluence of bursts of morphology III and IV than for bursts of MI and MII. The number of subcomponents of an injection impacts its detectability as well, since this is the main difference between bursts of MI/MII and MIII/MIV. The more subcomponents, the wider an injection's overall width, which, similar to scattering, might induce a large width mismatch as follows from Fig. \ref{fig:wth_dependence}. The MIV injections, unaffected by scattering, have wider overall widths because, on average, they have been injected with more subcomponents than the MIII injections. The derived difference between the pipeline's completeness for the intrinsic fluence of MIII and MIV bursts is therefore most likely an injection bias and not caused by an inefficiency of the pipeline.\newline

Aside from a fluence from which the HTRU-North SP search is complete, a more experimental fluence sensitivity limit can be obtained from the fluence recall curves in Fig. \ref{fig:fluence-recall}. Such a sensitivity limit reflects the brightness that SPs should minimally have for it to still be detectable. This corresponds to the fluence at which the pipeline's recall just gets larger than zero. Taking a recall of 0.05 for this limit and considering the recall curve for the unscattered bursts of morphology I, a fluence sensitivity limit is found of $\sim$0.16\,Jy\,ms. A similar value is acquired if the unscattered bursts of morphology I are taken and their fluences scaled as if they were injected with a width of 1\,ms. The injected SP with the lowest scaled fluence, which has still been detected, has this fluence. This experimentally obtained fluence sensitivity limit is $\sim$1.3 times higher than the theoretically derived limit from Sect. \ref{sec:search_sensitivity}. Interestingly, this is the same factor as found by \citet{ber2019}, concluding that the HTRU-North survey is 1.3 times less sensitive based on the re-detection of known pulsars through the periodicity search of the same HTRU-North data.

\subsubsection{Detectability discrepancies}\label{sec:D-discrepancies}
\begin{figure}
\centering
\includegraphics[width=0.5\textwidth]{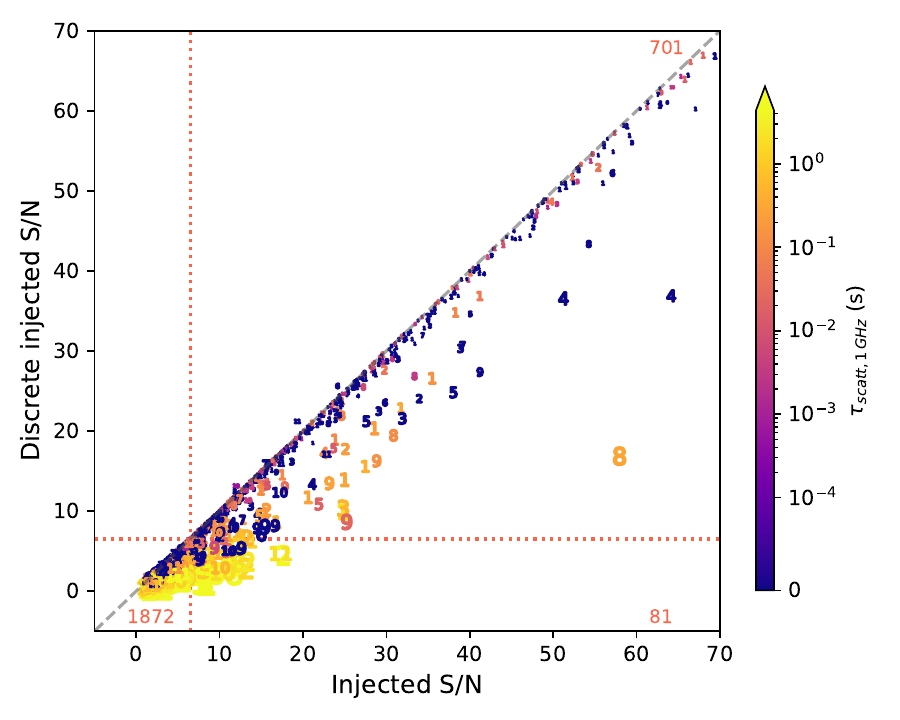}
   \caption{Examination of the effects on the detectability of SPs when searched for with a limited set of boxcar widths. The injected S/Ns of all injected SPs are plotted against their discrete injected S/N. The width difference between the best-fitting boxcar width and the actual propagated burst width of an SP is indicated by the plotted marker sizes. The scattering time (referenced to 1\,GHz) with which it is injected is represented by the marker colours and the marker symbols, which depict the number of subcomponents an SP is composed of. The dotted red lines indicate \heim's S/N search threshold of 6.5, and the number of SPs in each quadrant of the plot is given by the red numbers in the corners.}
   \label{fig:wth_dependence}
\end{figure}

While RFI makes it harder to detect SPs, the injection analysis was set out to explore circumstances under which the HTRU-North SP pipeline unexpectedly still detects SPs. These circumstances must exist, since Fig. \ref{fig:wth_dependence} indicates that 701 injections should be recoverable by \heim\ with an S/N above the set S/N threshold of 6.5. However, the SP pipeline actually detected 795 injections, meaning that a significant amount of injections were unexpectedly detected.

To investigate how these injections could have been detected, the injected S/Ns of all detected SPs are plotted against their recovered \heim\ S/Ns in Fig. \ref{fig:detected_injections}. As in Fig. \ref{fig:wth_dependence}, the colours of the markers represent the applied scattering times. The marker sizes depict the boxcar widths with which \heim\ recovered the injections, and lines are drawn from the discrete injected S/N of the injections (plotted in Fig. \ref{fig:wth_dependence}) to their recovered \heim\ S/N. These lines make it easier to see if injections have been recovered with a higher or lower S/N by \heim\ than expected. As indicated in this figure, 109 injections were detected even though their discrete injected S/Ns lie below \heim's detection threshold. Together with the 701 expected detections this makes a total of 810 possible detections. Thus, 15 injections expected to be detectable have also been missed by the pipeline in order to get to the 795 detections reported in Table \ref{tab:num_recoveries}.

As discussed in Sect. \ref{sec:detect&ident}, one of the missed injections was injected coincident with an RFI spike. The other 14 missed injections have been injected with an S/N close to the detection threshold. Since the recall curves in Fig. \ref{fig:snr-recall} attain a value close to 0.5 at the S/N detection threshold, the pipeline detects SPs around this threshold with an even probability. The 14 missed injections are therefore most likely missed by chance.

Several of the unexpected detected injections are injected with an S/N far from the pipeline's S/N detection threshold as seen in Fig. \ref{fig:detected_injections}. Their detection can thus not be attributed to chance and other reasons must exist that caused the pipeline to unexpectedly detect these injections. To determine these other reasons, from the 109 unexpected detections, SPs injected outside of the S/N range of five to eight have been manually investigated. Aside from chance, the following special circumstances can cause the HTRU-North SP pipeline to detect injections with a discrete injected S/N below the set S/N threshold:
\begin{enumerate}
    \item An SP is injected coincident with RFI and its S/N is boosted above the S/N threshold due to the added power of the RFI.
    \item When injected immediately next to (not on top of) a strong RFI spike, an SP can be injected with a much higher rms (and therefore S/N) than with what it is found after {\tt RFIbye} removes the RFI.
    \item If an SP is injected in channels that have an on average low rms while the full bandpass varies strongly, removal of the bandpass will cause the signal to be boosted in brightness.
    \item Integer conversion of a dim injection's profile can cause its morphology to lose complexity and its power to be concentrated in just a limited amount of data samples.
    \item The combination of over or under de-dispersion with downsampling in frequency and time can result in the aggregation of spread-out power into a few data samples aligned in time.
    \item Over-de-dispersion aligns down-drifted subcomponents\footnote{Morphological phenomena also known as the `sad trombone' effect.} in frequency over time, so their power is combined in the TS.
\end{enumerate}

From the unexpected detections 12 have an $S/N_{inj} > 8$. These bursts have almost all been injected with very high scattering times, as can be seen from Fig. \ref{fig:detected_injections}, causing their emission to be spread thinly over many data samples. The integer conversion of these bursts then deforms their morphologies such that their power is concentrated in the top part of their BW and so have higher recoverable S/Ns (see Appendix \ref{sec:A.int_inject}). The four bursts with no to small scattering timescales consist of many subcomponents causing their discrete injected S/N to drop below 6.5, but could still be detected due to reason six.

Twenty-two unexpected detected injections have an $S/N_{inj} < 5$, of which two injections are of morphology I, five of MII, four of MIII and the other eleven of MIV. This uneven distribution is caused by the above identified six unexpected detection circumstances predominately having an effect on the low S/N bursts of a specific morphology. While the two MI injections were detected due to a different reason, all but one of the MII injections were detected because of reason three. The MIII injections have most often been detected due to reason four, and reason six explains why half of the unexpected detected injection belong to MIV. SPs of MIV are not only the sole bursts that can be affected by circumstance six, their brightness can also be boosted by higher factors than most of the other circumstances. SPs with a (discrete) injected S/N as low as about two could therefore still be detected.

Circumstances two, three, and four are artefacts of the {\tt FRBfaker} injecting SPs in integer data and their occurrence rates are low. Circumstances one, five, and six can lead to the discovery of SPs with a lower intrinsic S/N than the detection threshold set by real blind searches for SPs. In particular, reason six is an efficient manner to boost an SP's recoverable power, as can be deduced from Figs. \ref{fig:snr-recall} and \ref{fig:fluence-recall}. For MIV bursts, the HTRU-North pipeline is complete for the highest fluence of $\sim$0.8\,Jy\,ms, while in terms of S/N it is complete for an S/N comparable to that of morphology I and II. This is because the time alignment of an MIV burst's subcomponents makes its morphology akin to that of an MI or MII burst, resulting in more of its power to be recovered and a higher detection S/N.

De-dispersion cannot only align MIV burst's subcomponents, in combination with downsampling it can also ensure the aggregation of smeared out power into several time aligned data samples. This enabled the detection of highly broadened SPs, such as the injections with a scattering time larger than the largest search width of \heim\ (see Fig. \ref{fig:detected_injections}). The combined use of de-dispersion and downsampling is thus a practical tool to correct dispersed signals and to (partially) mitigate the harmful effects of propagation and a low intrinsic brightness (in the case of MIV bursts) on the detectability of SPs.

\subsubsection{Performance FETCH models} \label{sec:fetch-models}
\begin{figure}
\centering
\includegraphics[width=0.5\textwidth]{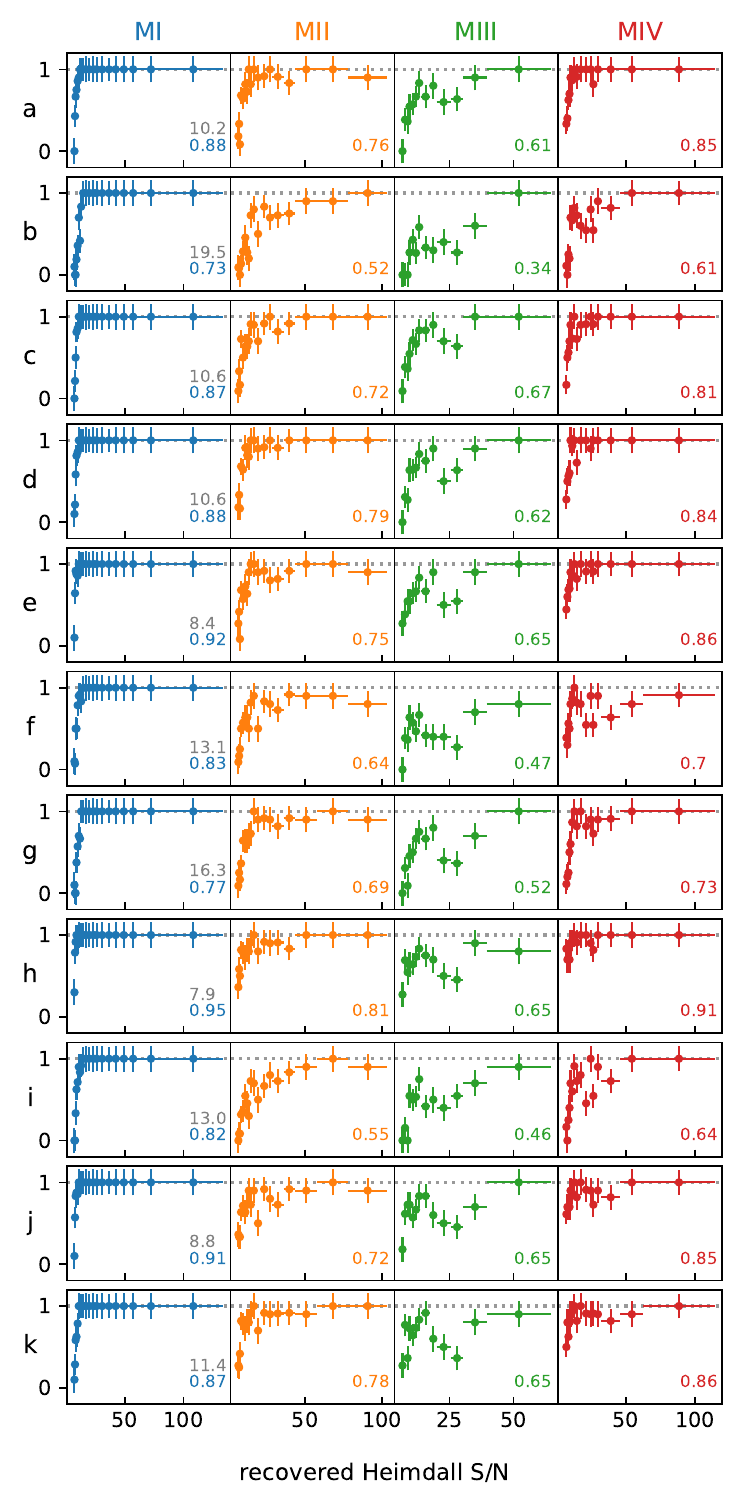}
   \caption{Recall curves for each of \textsc{fetch}'s classification models per FRB morphology as a function of the recovered S/N by \heim. The coloured numbers in the bottom-right corner of each subplot give the overall recall. For MI, the S/Ns for which the models attain a recall of 0.95 are also shown in grey, i.e. when they are complete.}
   \label{fig:fetch-recall}
\end{figure}

As described in Sect. \ref{sec:pipeline}, all candidates are evaluated by \textsc{fetch} and the letters of its models are stored, which returned a probability of $\geq$0.5 with which they deemed a candidate to be real. Since the applied models have not been re-trained, the stored models' letters of the detected injections can be used to inspect the native performance of these models on the HTRU-North data. The letters are therefore turned into recall curves for each of \textsc{fetch}'s models per morphology as a function of the recovered S/N by \heim\ and displayed in Fig. \ref{fig:fetch-recall}. The recall curves were determined in the same way as those in Fig. \ref{fig:snr-recall}. For low S/Ns, none of the models attained a recall value of 1, meaning that if \textsc{fetch} would have been blindly used to select SP candidates for the human assessment in Sect. \ref{sec:analysis}, more injections would have been missed.

The obtained overall recall values (coloured numbers in the subplots of Fig. \ref{fig:fetch-recall}) are much lower than the $>$0.99 reported by \citet{aab+2020} for all models. Also, only the models' recall curves for MI are well described by Eq. \ref{eq:exponential} and therefore have an S/N listed for which the models are considered complete (grey values). Since \citet{aab+2020} mentions that the models' performance is heavily influenced by the data with which they are trained, these findings are most probably due to differences between the injection data and the used training data. The latter consisted of SPs from RFI, pulsars, and simulated FRBs of morphology I and II that were injected with a maximum scattering time of 50\,ms and with a minimum S/N of 8. This explains the poor performance of all models for detected injections with a low S/N. The models have not been trained on such low-brightness or strongly scattered pulses. Especially since the low S/N detections often have larger scattering times (Fig. \ref{fig:detected_injections}) than the injections in \textsc{fetch}'s training data.

The performance of the models is also influenced by the way \textsc{fetch} determines candidates to be true positive events. It does this by evaluating the closeness of a candidate's de-dispersed dynamic spectrum to a straight line, and the closeness of its DM-time image to a bow-tie signature. The narrower the BW of a candidate, the more its de-dispersed dynamic spectrum shows a more point-like feature that might not be considered a line by \textsc{fetch}'s models. This might be why most wrongly classified injections have a smaller BW than the average of the detected injections, and why the models perform less well for MII than MI.

\textsc{fetch}'s wrongly classified MIII injections have subcomponents much further spaced apart than the average of the detected MIII injections. Their DM-time images will therefore show multiple (overlapping) bow-tie signatures in case their individual subcomponents are bright enough. This might confuse the models to then classify these injections as RFI. A hypothesis supported by the wrongly classified MIV injections showing smaller drift rates than the average applied rates for MIV's detected injections. These smaller drift rates make them look more similar to the MIII injections. A larger subcomponent separation does not increase mislabelling of the MIV injections, because in combination with a large drift rate, their subcomponents can be aligned in time through over-de-dispersion. Circumstance six from Sect. \ref{sec:D-discrepancies} thus ensures MIV injections to more clearly show a line in their de-dispersed dynamic spectra, even though their subcomponents might have a narrow BW, and a single bow-tie signature in their DM-time images. Due to circumstance six, MIV's recall values are the best after those of MI. Those of MIII are the worst, since the two effects described above both affect the injections of this morphology.

As no information about the accuracy\footnote{Accuracy reflects how often a model makes a correct classification, which includes the cases where RFI is correctly identified as not being a true SP (the true negative events)} of the models has been deduced, the recall values should be considered an indication of the performance of \textsc{fetch}'s models. Nevertheless, Fig. \ref{fig:fetch-recall} shows that the models produce different results for bursts of the four morphologies. In cases where SP searches rely on the classification of \textsc{fetch}, it is thus advisable to re-train the models on data that contain SPs with the morphologies searched for. If re-training is not among the options and the pre-trained models must be used, it might be advisable to never use a single model for the classification of SPs. Although, \citet{aab+2020} mentioned the use of just model $a$ as an option, using a combination of the results of models $a$, $c$, $e$, and $h$ will already improve \textsc{fetch}'s classification as these show, for each of the morphologies, recall curves most in line with Eq. \ref{eq:exponential} and have high overall recall values.

\section{Results} \label{sec:results}
To perform the just described injection analysis, the first 1500 HTRU-North PTs have been processed and searched for SPs. Alongside a notion of the efficiency of the pipeline, the following results are obtained through the analysis of these PTs.

\subsection{RFI environment at Effelsberg} \label{sec:RFI-env}
As mentioned in Sect. \ref{sec:pipeline}, \heim\ is run twice in the HTRU-North SP pipeline, once before and once after the data are cleaned by {\tt RFIbye}, and for both instances the valid candidate counts of an inspected beam file are recorded. Taking the median of the valid candidate counts of the beam files of a PT results in the median valid candidate count per PT. This count is agnostic about the number of beam files recorded for a PT or any receiver-induced RFI that only affects a single beam. Therefore, the median valid candidate count can be used to investigate the RFI background in which the HTRU-North observations are performed.

However, a search for SPs in TSs, with $N_{s}$ samples and through the convolution of boxcars with widths of $2^{j}$ samples, is always expected to find some candidates due to instrumental noise alone. \citet{cm2003} derived that the expected number of SP candidates, $N$, found above a set S/N threshold and in $N_{DM}$ DM trials due to instrumental noise is given by
\begin{equation}
    N\textrm{(>\,S/N)} = N_{s}N_{DM}P\textrm{(>\,S/N)}\sum_{j=0}^{j_{max}} 2^{-j},
\end{equation}
where $P$(>\,S/N) is the integrated Gaussian probability from the S/N threshold to infinity. For the HTRU-North SP pipeline, this yields a value of $N$(>\,6.5) $\simeq$ 0.6 per beam file of a PT. The number of times this happens in exactly three adjacent boxcar search or DM trials is even lower. An identified valid candidate is thus less likely to originate from instrumental noise, with a likelihood that is reduced the more grouped members are required for a candidate to be considered valid.

\begin{figure*}
    \centering
    \includegraphics[width=\textwidth]{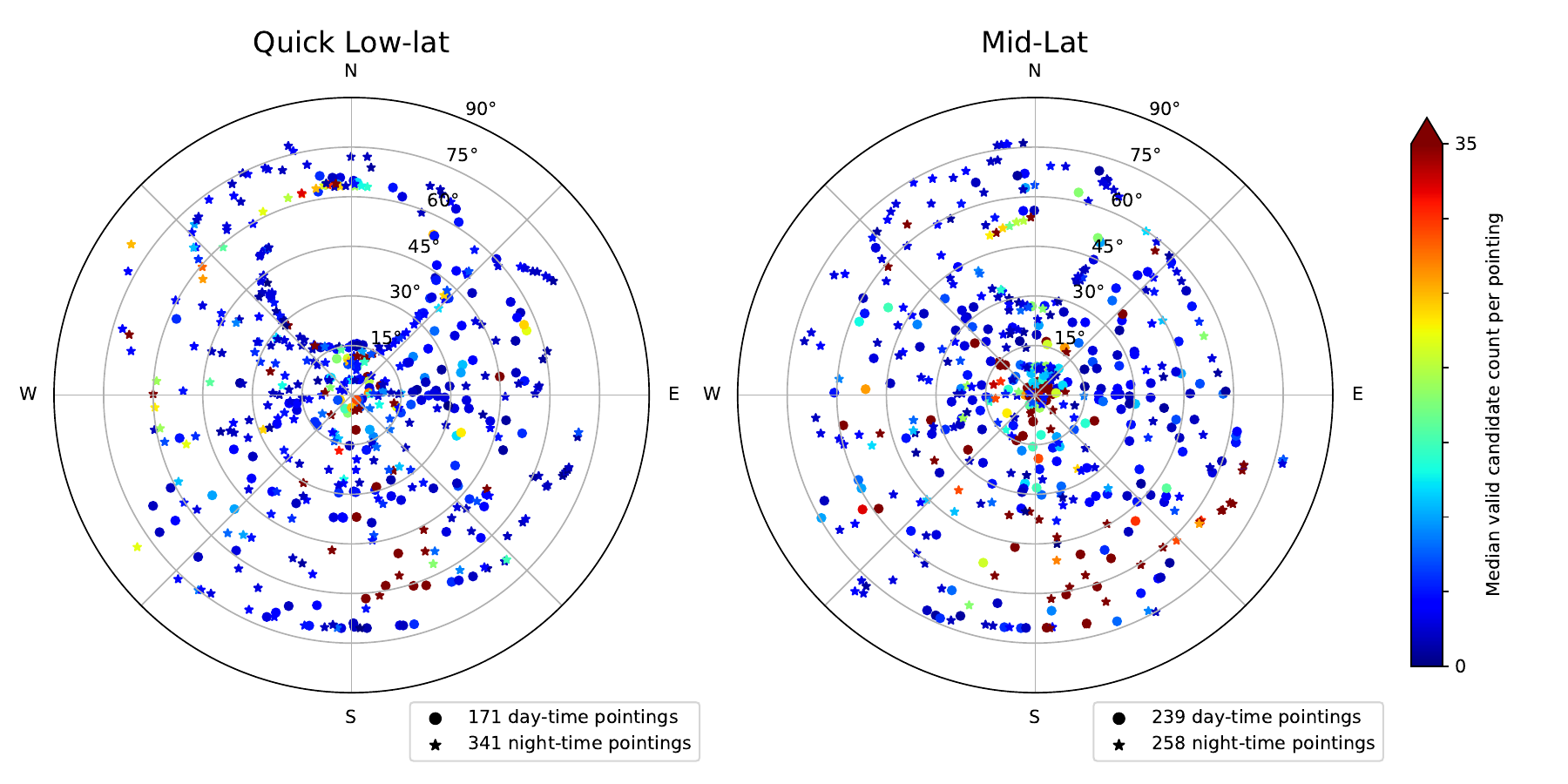}
    \caption{Distribution of selected quick low-lat and mid-lat PTs over azimuth and elevation. PTs recorded during the day are indicated with a circle, and those recorded at night are represented with a star. The colours show the median of the number of \heim\ candidates per PT before RFI excision. These values are an indication of the amount of RFI recorded per PT. Note, the upper end of the candidate count scale is truncated at the 0.9 quantile of the PTs' candidate count, meaning that some `dark red' PTs contain a factor of 10 to 100 more candidates than most PTs.}
    \label{fig:rfi-skymap}
\end{figure*}

In Fig. \ref{fig:rfi-skymap}, the median valid candidate counts, obtained before running {\tt RFIbye}, are plotted per PT at the observed celestial sky location of the PT. The majority of the PTs have at least a median valid candidate count of four. Since candidates originating from known sources are not included in this count, the HTRU-North PTs are thus contaminated with RFI. Though the rate of contamination varies strongly across the PTs, the counts shown in Fig. \ref{fig:rfi-skymap} are truncated at the 0.9 quantile of the counts per PT. From this figure, it is evident that the amount of observed RFI depends on the direction in which Effelsberg is oriented. The telescope is built in a valley to shield it from RFI. However, when it is pointing close to zenith, its receiver cabin is raised above the hills, making it more susceptible to RFI, explaining the higher candidate counts around zenith. The higher median valid candidate counts for PTs in the south--south-east direction of the telescope can be explained by the valley opening in that direction that provides less shielding. The observed RFI environment is thus expected, but it increases the number of false positive detections for certain telescope orientations. Hence, the use of {\tt RFIbye} to reduce this false positive rate.

For the second run of \heim, 96\% of the processed PTs have a median valid candidate count below four, indicating a significant reduction in the amount of RFI present in all the PTs. 77\% of the PTs even attained a count of zero, which is consistent with the expected false positive rate. Among these PTs are PTs that even have an initial median valid candidate count above the truncation threshold. Thus, {\tt RFIbye} performs really well in removing RFI from the HTRU-North data. It is, however, less successful in removing RFI from highly contaminated beams in which the data's statistics are skewed by the presence of RFI. The beam files of the 4\% of PTs with a median valid candidate count greater than or equal to four after the application of {\tt RFIbye} already contained a lot of RFI before they were cleaned. In the future, {\tt RFIbye} could be updated to better calculate data statistics in the presence of strong RFI and thus be able to more adequately clean these data as well.

Together with the result of Sect. \ref{sec:detect&ident} that no injected SPs have been removed by {\tt RFIbye}, it is therefore worthwhile to keep utilising {\tt RFIbye} in the HTRU-North SP pipeline. It improved the RFI situation significantly, as seen from the valid median candidate counts before and after running {\tt RFIbye} whilst leaving desired signals in the data. To assure the best RFI mitigation with {\tt RFIbye}, it is good to monitor its performance over time in case the RFI environment at Effelsberg changes, and a different set of threshold values might result in cleaner data.

\subsection{Known pulsars and RRATs} \label{sec:re-detections}
In the beam files of the (pre-)analysis PTs, a large number of SPs were detected at a Galactic location and around a DM corresponding to a total of 22 known sources listed in the Australia Telescope National Facility Pulsar Catalogue (\psrcat)\footnote{\url{https://www.atnf.csiro.au/research/pulsar/psrcat}} \citep{mhth2005} and RRATalog\footnote{\url{https://rratalog.github.io/rratalog/}}. Folding these beam files with the DM and period of the pulsars, as listed in the \psrcat, resulted in the clear re-detection of 19 known pulsars. Table \ref{tab:re-detections} lists all re-detected known sources.

\begin{table}
\caption{Known pulsar and RRAT re-detections within the first 1500 searched HTRU-North PTs.}
\label{tab:re-detections}
\centering
\begin{tabular}{l c l | c}
    \hline
    \noalign{\smallskip}
    & Pulsars & \multicolumn{1}{c}{} & RRATs \\
    \noalign{\smallskip}
    \hline
    \noalign{\smallskip}
    B0037+56 & B1911-04 & B2035+36   & \\
    B0052+51 & B1920+21 & B2148+63   & \\
    B0059+65 & B1935+25 & J0212+5222 & \\
    B0105+65 & B1937+24 & J0631+1036 & J1819-1458 \\
    B0540+23 & B2000+40 & J1844+00   & \\
    B1821-19 & B2011+38 & J1852-0635 & \\
    B1834-10 & B2021+51 & J2352+65   & \\

    \noalign{\smallskip}
    \hline
\end{tabular}
\end{table}

One of the folded re-detections includes B1937+24 with an S/N of $\simeq$ 27 of its folded profile, though, for which only one SP was found (S/N$_{SP} = 10.3$). This detection shows the existence of clear emission at L-band, even though the \psrcat\ does not contain a mean flux density entry at 1.4\,GHz for this pulsar. Pulsar B0052+51 is one of the three sources that could not be detected in their folded data, which is unexpected since its estimated folded S/N at L-band, based on its \psrcat\ entries, is about 165. \citet{wes2006} identified this pulsar to have a large modulation index that suggests the power of its SPs to vary strongly from pulse to pulse. Together with it being detected in a side lobe, this might explain why seven evident SPs were found of B0052+51 while no average emission was detected.

The other two sources not detected with a folded profile of significance (S/N > 6.5), and for which only a single but convincing SP was observed, are pulsar J2352+65 (S/N$_{SP} = 8.1$) and RRAT J1819-1458 (S/N$_{SP} = 9.6$). Their long periods, 1.16 and 4.26\,s respectively, probably result in them having emitted too few observable SPs during their observation to enable their detection with a folded profile.\newline

To determine the true sensitivity of the HTRU-North survey, \citet{ber2019} selected $\sim$200 known pulsars that should be detectable in the available mid-lat PTs based on their \psrcat\ entries and periodicity searched the corresponding PTs for these pulsars. From the searched PTs in which a known pulsar was found, 16 PTs overlap with those searched here for SPs. In eight of these the known pulsars were also re-detected through the detection of their SPs. Half of the known pulsars selected by \citet{ber2019} and present in the analysed PTs thus emit SPs bright enough to be individually detectable, which is slightly more than the fraction found by \citet{bb2010} for the HTRU-South survey ($\sim$0.4).

Reasons for why the other 14 pulsars detected through their SPs are not in the selection list of \citet{ber2019} (albeit them being bright) are as follows. Three pulsars do not have an L-band flux density entry in the \psrcat\ and are therefore not included in the selection list. Or, they were not yet included in the \psrcat\ at the time the selection list was made, which is the case for J0212+5222 as it is one of the pulsars first detected through periodicity searching the HTRU-North data \citep{bck+2013}. Most other pulsars are detected in a side-lobe of the receiver and, hence, found in a different PT than in which they were expected to be found based on their sky locations. These pulsars therefore did not end up in the selection list of \citet{ber2019}.

\subsection{Miscellaneous}\label{sec:cand-statistics}
\begin{figure*}[!ht]
    \centering
    \includegraphics[width=\textwidth]{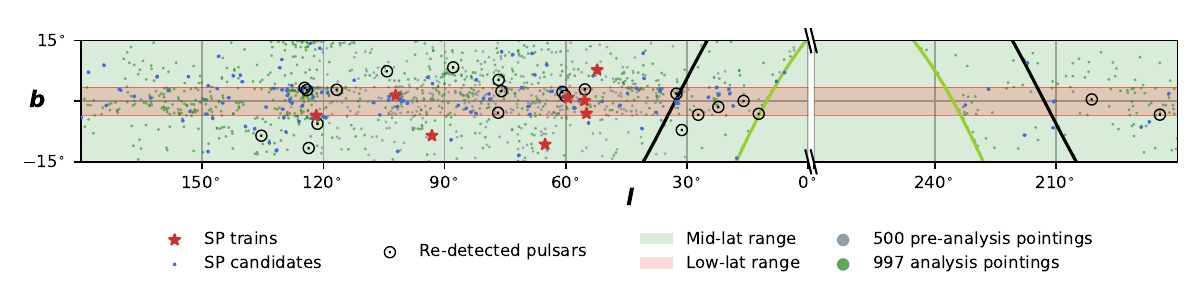}
    \caption{Galactic positions of the (re-)detected pulsars, SP trains, and SP candidates. The locations of the selected (pre-)analysis PTs are also shown with a diameter equal to that of the Effelsberg 21\,cm multibeam receiver's beam pattern on the sky. The black line indicates the equator, and the green line is the declination limit of Effelsberg.}
    \label{fig:pointing-skymap}
\end{figure*}

After RFI excision, 30\,280 valid candidates were obtained from the processed analysis PTs. These do not include SPs from known sources and have been manually inspected using the custom build \heim\ candidate plotter (Appendix \ref{sec:cand_viewer}). From the plotter's diagnostic plots and \textsc{fetch}'s classification scores, 507 candidates were deemed potentially real and their properties were recorded in the HTRU-North SP SQLite database.

Among the 507 candidates, several stand out because they have been detected in a single beam, with similar DMs. They so seem to form a train of SPs aligned around a specific DM, akin to the SPs of known pulsars. These specific DMs are all smaller than 1.5 times the DM contributed by the Galaxy\footnote{A cut-off often quoted for FRB SPs, to consider them as extra-galactic (see for instance \cite{kbj+2018}).}, estimated with the NE2001 model \citep{cl2002} for the SP trains' Galactic locations. For the eight SP trains, a tentative period could also be found that can divide the time in between their SP arrival times a whole number of times with a one-hundredths precision. They might therefore originate from yet unknown neutron stars. Table \ref{tab:pulse_trains} presents all the details of the detected eight SP trains. Non-valid candidates are included for some of the SP trains, since they showed a similar morphology as the valid candidates of that specific train, and to improve the determined period.

Subtracting the SP trains' SPs from the 507 identified candidates leaves 483 detected SP candidates with an S/N $\lesssim$ 8. If these SP candidates are real, they are produced by faint sources. To reduce the chance of them originating from RFI, only those SP candidates found in PTs stripped of RFI by {\tt RFIbye} (i.e. with a median valid candidate of zero) are further taken into account. With the additional requirement that an SP candidate must consist of at least five members, the chance of it arising from instrumental noise is further reduced as well. 141 SP candidates are so identified that might have a true astrophysical origin. These SP candidates are listed in Table \ref{tab:sp_candidates} and their Galactic positions, together with those of the detected SP trains and known pulsars, are shown in Fig. \ref{fig:pointing-skymap}. Interestingly, more SP candidates seem to be detected in the quick low-lat PTs (85) than in the mid-lat PTs (56). Since the total observation time of the quick low-lat PTs and the mid-lat PTs in which the SP candidates are found is the same, this tentatively hints towards a latitude dependence on the detection rate of faint SPs. The interpretation of this dependence shall be addressed elsewhere if it persists when more PTs have been analysed.

\section{Discussion} \label{sec:discussion}
In Sect. \ref{sec:analysis} an injection analysis was performed that mimicked the full search for SPs in the HTRU-North data and yielded the results presented in Sect. \ref{sec:results}. From these efforts, the following insights are obtained.

{\tt RFIbye} reduces the amount of RFI considerably and may do so better than other existing RFI mitigation tools. However, it has difficulties with the calculation of data statistics whenever there is a large amount of RFI present in the data, which reduces its ability to remove RFI. Improving data statistics calculation in the presence of strong RFI would allow {\tt RFIbye} to better clean this data and \heim\ to report fewer false positive candidates.

The median valid candidate count, introduced in Sect. \ref{sec:RFI-env}, is in effect a coarse coincidence filter that can be applied to any receiver with multiple beams. If such a count is obtained before RFI mitigation is applied, it can be used as an indirect measure of RFI contamination. In the case of the 21\,cm multibeam receiver of the 100\,m Effelsberg Radio Telescope, the median valid candidate count could be used to show the RFI contamination of its PTs to be directional dependent. If the telescope is pointed close to zenith or in a south--south-easterly direction, it is much more susceptible to RFI.

\heim\ was shown to only search for SPs up to where data still contain the entire dispersion sweep across the data's frequency range for the highest searched DM. In the case of the HTRU-North SP pipeline, this means that the last approximately 5\,s of each beam file are not searched for SPs. Any SP present within these 5\,s will therefore not be detected, regardless of the DM of the SP or how well the data was cleaned with {\tt RFIbye}. For the PTs analysed, this means that over two hours worth of data have not been searched for SPs. Hence, it is desirable to append fake data to the end of a PT's beam file before step 2 in the pipeline (Sect. \ref{sec:pipeline}), as has been done by \citet{myw+2022}. This ensures that \heim\ actually processes all the available data and detects SPs near the end of a beam file. This assumes that the S/N of these SPs is not too far reduced by the possibility that their dispersion sweep is not entirely contained within the unpadded data.

More SPs can be found by allowing \heim\ to search with wider boxcar widths than is currently done by the HTRU-North SP pipeline. A high width discrepancy between the widest boxcar used for the SP search and the true (propagated) burst width might render the burst undetectable (Sect. \ref{sec:pipeline-recall}). If \heim\ would have been set to search for much wider bursts than the current maximum of 13.98\,ms, 81 of the injected SPs might have been additionally detected as follows from \mbox{Fig. \ref{fig:wth_dependence}}.

Furthermore, \heim's performed de-dispersion and downsampling caused some injections to be unexpectedly detected (Fig. \ref{fig:detected_injections} and Sect. \ref{sec:pipeline-recall}) as well. Over-de-dispersion can time align faint subcomponents of an MIV's burst, causing its morphology to resemble that of MI or MII, and so render its frequency-averaged pulse profile detectable. Conversely, this might also mean that faint FRB detections with an apparent morphology I or II actually consist of multiple over-de-dispersed subcomponents. If morphology IV indeed describes FRB repeat bursts \citep{pgk+2021,cgsy2022}, it could be interesting to re-observe the sky locations of faint one-off FRBs, as it is possible that they are FRB repeat bursts falsely identified as MI or MII bursts due to over-de-dispersion.

Allowing \heim\ to search for wider bursts would increase the number of detected SPs and therefore also the number of false positives produced by the pipeline. A reliable manner is needed to keep the total number of SP candidates, to be manually inspected, to a human-manageable rate. \textsc{fetch} can potentially be used for this if its classification models are re-trained on the HTRU-North data containing a balanced set of SPs originating from RFI, pulsars, and FRBs. An HTRU-North training dataset containing FRBs can be obtained with the use of the {\tt FRBfaker}. SPs of the four different FRB morphologies and with low S/Ns can then be injected, improving \textsc{fetch}'s overall performance. Its current models will miss such signals, because they have not been trained to find them (Sect. \ref{sec:fetch-models}).

Re-training of \textsc{fetch}'s models will improve its capability of correctly identifying found SPs by \heim, but as pointed out in Sect. \ref{sec:fetch-models}, the way in which \textsc{fetch} performs its classification might also hamper the reliability of this classification. Narrow-band SPs or ones consisting of multiple subcomponents not drifting down in frequency over time are particularly affected. Implementation of a sub-banded search as applied by \citet{tpp+2024} could minimise these effects. They effectively applied a sub-banded search on the HTRU-South data, with a similar BW as the HTRU-North data, and so discovered another 18 FRBs in this data. The additional benefit of such a sub-banded search is that it would also improve the effectiveness of the pipeline as a whole to detect band-limited SPs.

With the above improvements applied, the HTRU-North SP pipeline is expected to find even more SP candidates. Even without, extrapolating the found 141 SP candidates and eight SP trains to the entire available mid-lat data, the pipeline is expected to find approximately 4000 and 225 of them, respectively. That is, if the potential detection rate dependence on Galactic latitude (Sect. \ref{sec:cand-statistics}) does not persist. With this many potential sources, it is likely that at least a few new neutron stars and FRBs will be detected with the continuation of the analysis of the HTRU-North data.

\section{Conclusions} \label{sec:conclusions}
In this paper, we have presented a new and dedicated SP pipeline aimed at searching the available HTRU-North data for SPs. In its current state, it is capable of detecting SPs of known sources (Sect. \ref{sec:re-detections}), which emphasises the importance of SP searches, as some sources were not detected when periodicity searching the same data. We identified eight SP trains that might originate from yet undetected neutron stars and 141 faint isolated SP candidates of unknown origin (Sect. \ref{sec:cand-statistics}). Follow-up research is needed to investigate their authenticity and origin. 

Through injection tests with SPs of varying morphologies, the pipeline was shown to be complete for SPs with an \mbox{S/N $\gtrsim$ 11} and to have a fluence sensitivity limit of $\sim$0.16\,Jy\,ms (Sect. \ref{sec:pipeline-recall}). Sect. \ref{sec:D-discrepancies} established that in some cases, faint SPs or SPs with a wider width than searched for can be detected due to de-dispersion and downsampling of the data during the search. If the detection of SPs relies on the classification performance of \textsc{fetch}, its models are best re-trained on the user-specific data, as we discussed in Sect. \ref{sec:fetch-models}. The {\tt FRBfaker} can help in creating training data for this, as it can (re-)create and inject SPs of highly complex morphologies in filterbank data.

New observations with Effelsberg's 21\,cm multibeam receiver are advised to take the directional dependence on its susceptibility to RFI into account, as identified in Sect. \ref{sec:RFI-env}. Such newly obtained data might be less affected by RFI. Nonetheless, in case the data are affected by RFI, {\tt RFIbye} can be used to effectively remove it. For the HTRU-North data, this tool was capable of significantly reducing the amount of RFI in 96\% of the processed PTs, and 77\% of them could even be considered free from RFI. The custom codes presented here thus form powerful additions to the available SP processing toolkits.

\section*{Data availability}
The full version of Table \ref{tab:sp_candidates} is available in electronic form at the CDS via \url{https://cdsarc.cds.unistra.fr/viz-bin/cat/J/A+A/707/A10}.

\vfill
\mbox{}

\begin{acknowledgements}
Based on observations with the 100-m telescope of the MPIfR (Max-Planck-Institut f\"ur Radioastronomie) at Effelsberg. The authors wish to thank all the observers who used the telescope to help accumulate the HTRU-North data over the past years, and so enabled the research presented here. Also, we would like to thank the referee for the effort of providing us with detailed and highly useful comments.
\end{acknowledgements}

\bibliographystyle{aa}
\bibliography{master}

\begin{appendix}
\onecolumn

\section{Heimdall candidate viewer}\label{sec:cand_viewer}
\begin{figure*}[h!]
    \centering
    \includegraphics[width=0.63\textwidth]{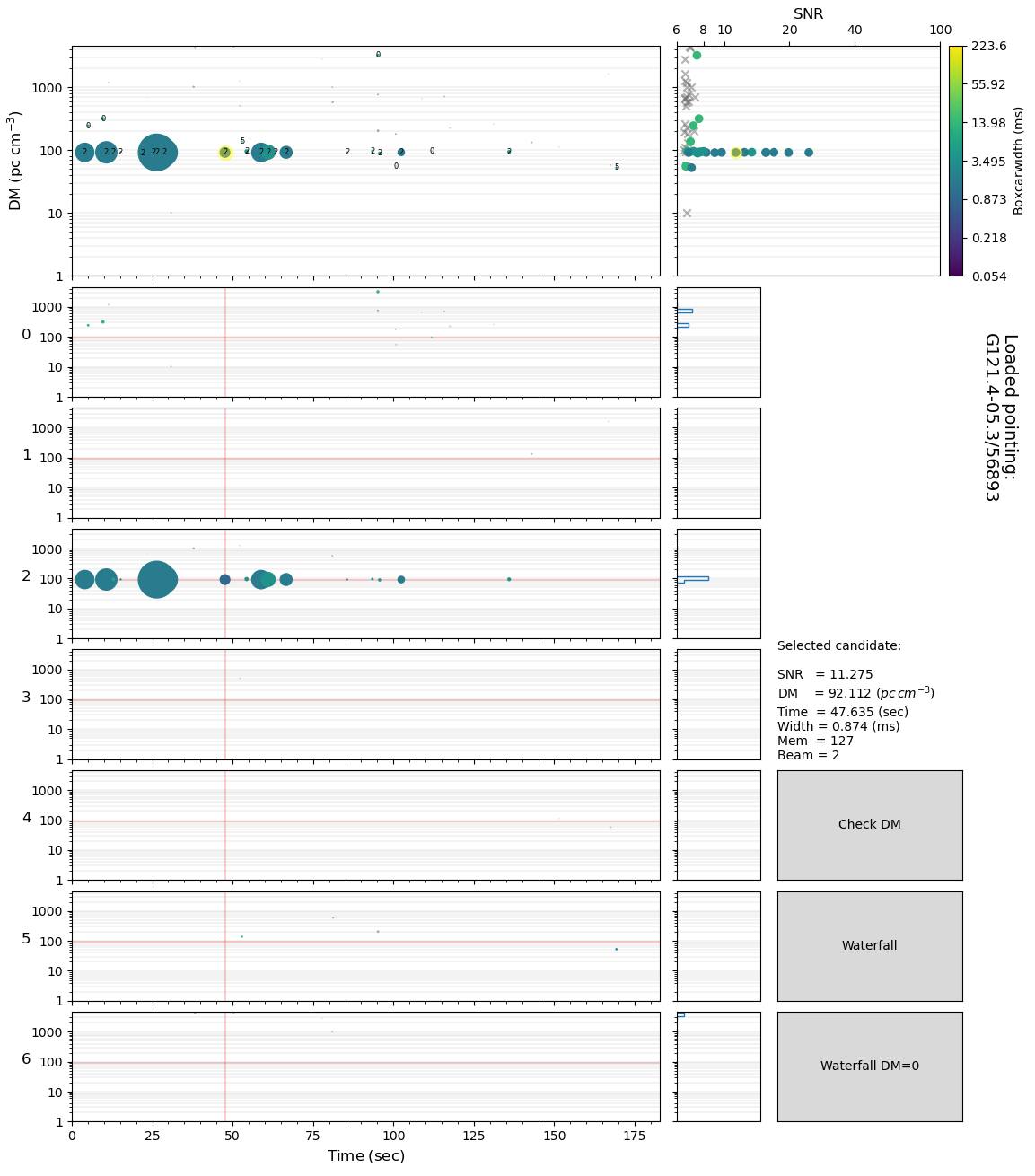}
    \caption{Candidate overview plot produced by the custom candidate viewer showing the candidates found of pulsar B0037+56. As can be seen, this pulsar is detected in beam two of Effelsberg's 21\,cm receiver and details about one of the detected SPs are listed on the right of the figure. The red crosshairs indicate the selected SP's time-DM position in all time-DM panels for the seven beams of the analysed PT. These make it easier to visually identify possible candidates coincident in time in other beams, and candidates belonging to an SP train in the same beam.}
    \label{fig:cand_overview_plot}
\end{figure*}

\vspace{-1.1em}
\begin{multicols}{2}\noindent
For the HTRU-North SP pipeline, an interactive candidate plotting tool\footnote{\url{https://gitlab.com/houben.ljm/heim_cand_plotter}} was developed to enable the quick and manual inspection of \heim\ candidates resulting from the fifth step of the pipeline (Sect. \ref{sec:pipeline}). The tool, inspired by \heim's {\tt trans\_gen\_overview.py}, is built up from DM-time plots, one for each beam of an analysed PT, that show the found candidates in a beam (Fig. \ref{fig:cand_overview_plot}). The candidate's marker type, size, and colour give information about the candidate's origin (possible RFI type if applicable), S/N, and width, respectively. The DM-time plots are aligned in time, so any candidate, coincident in time with candidates in other beams, can be spotted in an instance. As a confirmation of a possible coincidence of candidates, another DM-time plot is included, which forms the union of all candidates of all beams in which coincident candidates are marked as RFI. Determination of the coincidence of candidates across the beams of a PT is done with \heim's {\tt coincidencer}. 

A histogram is shown next to the DM-time plots of individual beams that shows the number of candidates in a specific DM bin. These histogram plots can ease the identification of the presence of multiple pulses with a similar DM that might originate from pulsars or RRATs. A DM-S/N plot is shown next to the DM-time plot of the coincidenced candidates, which is useful to get a more accurate idea about the S/Ns of candidates than can be obtained from their marker sizes. Especially since markers are given the same size for candidates with an S/N of 100 or higher. Together, these plots provide an overview of all candidates and their main characteristics, found in the beam files of a PT. From these plots, specific events, such as RFI storms and back-end power losses, can easily be recognised, since they cause very distinct patterns in the combination of plots shown.

If a plotted candidate looks real, it can be inspected in more detail by clicking on its corresponding marker in any of the plots. This will show all its properties as determined by \heim, and enables the creation of an array of dynamic spectra of the candidate either for the DM at which it is detected, or a DM of zero. The array of dynamic spectra is, along its rows and columns, downsampled over frequency and time, respectively, to make the presence of a possible signal more apparent. With a candidate selected, a DM-time image of the candidate can be created. These detailed diagnostic plots help the validation of detected \heim\ candidates.
\end{multicols}

\vspace{-1.1em}

\section{Drawn observational parameters for the injected synthetic SPs}\label{sec:A.inj_params}
\vspace{-1.1em}
\begin{multicols}{2}\noindent
Prior to injecting synthetic SPs in the HTRU-North analysis PTs, an SQLite database was filled with observational parameters that define the shape of the injections' dynamic spectra, as described in Sect. \ref{sec:inj_param_space}. This was done in the following manner.

Each synthetic SP was first assigned a `fluence' of its pulse profile at which it should be injected. These frequency-averaged fluences were drawn from a power-law distribution from 0.02 to 6.1\,Jy\,ms with a spectral index of -1.5 \citep{msb+2019}. The lower limit was chosen below the theoretical sensitivity limit to investigate potential circumstances under which the pipeline unexpectedly detects such low SPs. Then, each burst was given a `DM' randomly drawn from an exponentially modified Gaussian distribution fitted to the DM distribution of known FRBs and assured to lie between 1.5 and 5.000\dm. The intrinsic full width at half maximum (FWHM) of bursts or burst subcomponents (for FRB morphologies III and IV), `subC\_width', was taken from a log-uniform distribution of the sampling time of the used receiver of the Effelsberg Radio Telescope (54.61\,\textmu s) to 10\,ms. Next, all bursts were assigned a scattering time, `$\tau_{scatt}$', from a log-uniform distribution between 8.56e$^{-7}$ and 10.11\,s, independent of their selected DM. The latter was done to simulate over- and under-scattered bursts. The $\tau_{scatt}$ limits arise from the minimum and maximum scattering times, contributed by the Galaxy and estimated with the NE2001 model \citep{cl2002}, for lines of sight within the low and mid-lat ranges and referenced to 1\,GHz. Whenever $\tau_{scatt}$ is smaller than 20\% of the SP's dispersion broadened width at 1.4\,GHz, the scattering time was set to zero. With the DM, subC\_width and $\tau_{scatt}$ an injection time was determined to completely inject the burst and its dispersion delay at a random time in a PT's beam file.

As morphology I describes bursts consisting of a single subcomponent that span the full BW, these bursts are fully described with the observational parameters now determined. For the other morphologies, bursts were assigned a `BW' with a FWHM randomly drawn from a log-uniform distribution from 10\,MHz to the full BW. The reference frequencies, `$f_{ref}$', specifying the centre of the Gaussian envelopes, were chosen from a uniform distribution such that the bursts get injected within the band of the 21\,cm multibeam receiver. For morphology III and IV, consisting of multiple subcomponents, the number of subcomponents, `$N_{comp}$', that they are built up from are taken from a log-uniform distribution from 2 to 12 components. The time between these subcomponents, `subC\_time', was also drawn from a log-uniform distribution from the subcomponents' width to their width times the number of subcomponents. Lastly, the most characteristic aspect of FRB repeat bursts, the gradual displacement of their subcomponents in frequency over time, is described by the `drift rate'. Drift rates for pulses belonging to the fourth FRB morphology were randomly drawn from a uniform distribution from BW$_{chan}$/subC\_time to the full BW divided by the total width of the bursts (i.e. subC\_time $\cdot\ N_{comp}$). With the selection of drift rates, the bursts' $f_{ref}$ and BW were taken into account to ensure that subcomponents do not fully drift outside of Effelsberg's frequency band.\newline

\vspace{2cm}

\begin{figure}[H]
\centering
\includegraphics[width=0.5\textwidth]{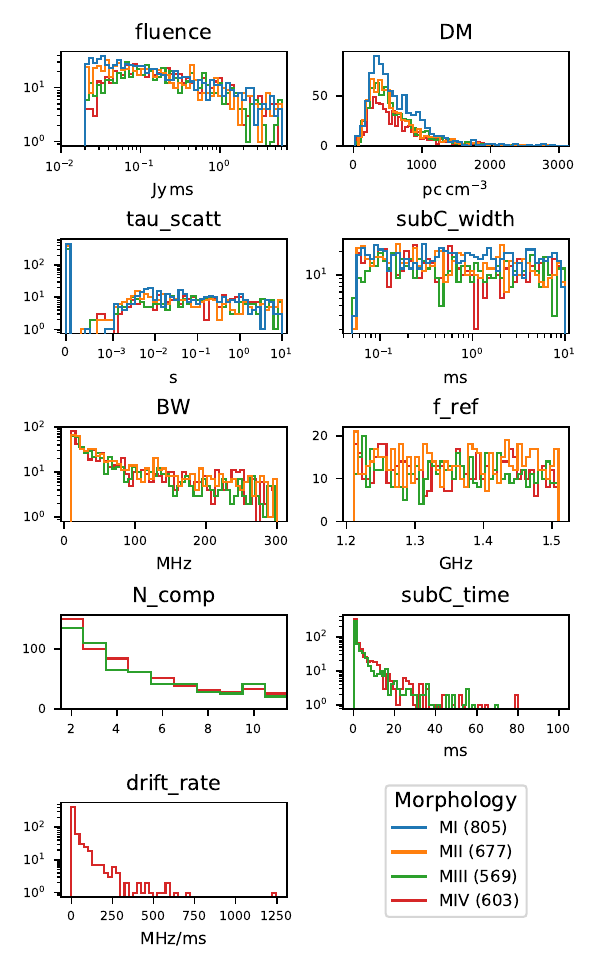}
   \caption{Histograms showing the distributions of the drawn parameters that were used for the injections with an $S/N_{inj} > 1$ and injected within \heim's search range. Table \ref{tab:injectionparams} lists the theoretical distributions and their limits from which these parameters were drawn. The number of injections above the applied $S/N_{inj}$ threshold and within \heim's search range are, per morphology, given in the legend behind their corresponding morphology name.}
   \label{fig:inj_params}
\end{figure}

\vfill
\mbox{}

\end{multicols}

\newpage
\section{FRBfaker performance tests} \label{sec:A.faker-tests}
\vspace{-1.1em}
\begin{multicols}{2}\noindent
{\tt FRBfaker} is made publicly available for others to inject synthetic SPs in their data, as has been done by \citet{myw+2022,lmz+2023,pbp+2024,gls+2025,lsm+2025}. It is therefore designed as a general toolkit, whose operation can easily be tailored to accommodate a user's needs. To ensure these needs are met, three final tests were carried out alongside the many intermediate tests performed during its development. With the first test, the {\tt FRBfaker}'s S/N scaling is investigated, the second test examines the effects of zero-DM filtering and bandpass removal on the injections' resultant brightness, and with the third, its ability to create complex burst morphologies is explored.

\subsection{S/N scaling test}\label{sec:A.scaling_test}
Regardless of the requested scaling of an injection's brightness, by providing a desired fluence or S/N, the {\tt FRBfaker} will use an S/N value to inject it with an appropriate power based on the rms of the data's noise. The `original {\tt FRBfaker} scaling' method, as described in Sect. \ref{sec:FRBfaker}, makes use of the radiometer equation (Eq. \ref{eq:radiometer}) to calculate the area an injection's pulse profile should have to inject it with a desired fluence or S/N in its TS. To determine how well the calculated area reflects the desired brightness of the injection, the following test is performed.

One hundred fake 8-bit filterbank files are created with the same time and frequency resolution as the HTRU-North data and filled with pure Gaussian noise. In each of these files, ten DM-smeared SPs are injected of morphology I with the same S/N, a DM of 312.526\dm, and an intrinsic width of 0.874\,ms. The DM and width are chosen such that they coincide with an applied search trial DM and boxcar width of \heim. Therefore, its reported pulse parameters can be considered to have the most optimal recoverable values, and deviations from the injected values are due to injection inaccuracies by the {\tt FRBfaker}. The bursts are injected per file with a requested S/N randomly drawn from a log-normal distribution from an S/N of eight to eighty.

\begin{figure}[H]
\centering
\includegraphics[width=0.48\textwidth]{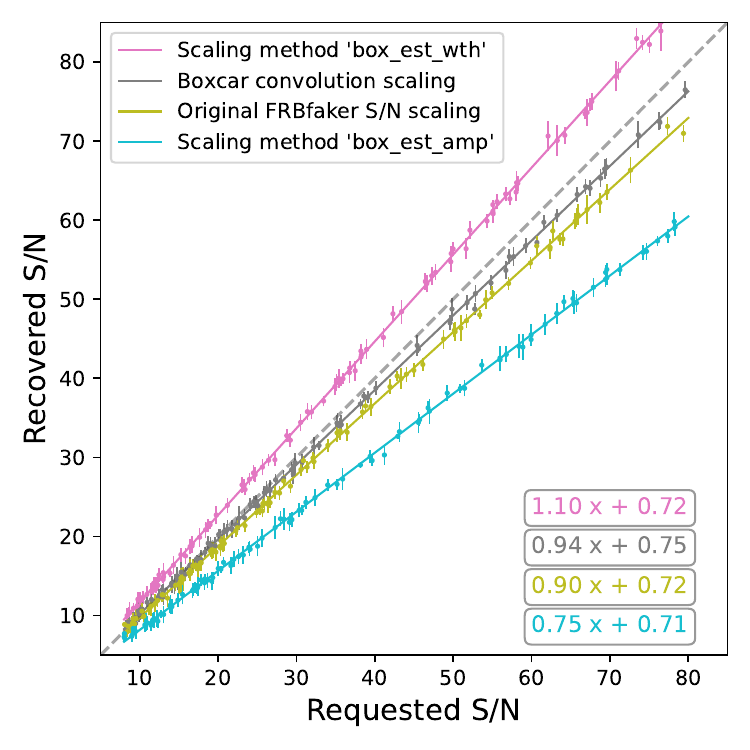}
   \caption{Recoverability of S/N for SPs injected with the {\tt FRBfaker} for two different methods of the S/N calculation (Original and Furby equivalent scaling) and three distinct ways of determining the injected SP properties (boxcar convolution, boxcar estimation based on width, and on the pulse's maximum value). The mean value at which \heim\ detected the ten bursts per file is plotted together with the spread in these values. A one-to-one S/N relation is indicated with the dashed grey line.}
   \label{fig:snr_recovery}
\end{figure}

From Fig. \ref{fig:snr_recovery}, the original {\tt FRBfaker} scaling curve, it follows that under these ideal circumstances \heim\ recovers the bursts with an on average $\sim$10\% lower S/N than their requested S/N values. Even though this discrepancy is of an expected order for bursts randomly injected in Gaussian noise, this scaling method assumes that Gaussian-shaped bursts are injected. The discrepancy might therefore be larger whenever this is not the case, for instance, when bursts of morphology III and IV are injected. A scaling method more agnostic to the morphology of the injections would thus increase the general applicability of the code.

Hence, the S/N scaling as applied by {\tt Furby}\footnote{\url{https://github.com/vivgastro/Furby}} is investigated. Its method is more data driven and scales the injection's area such that it has a similar area as a boxcar of one time-sample in width, and with the requested S/N. Since the profile's area is determined through the convolution of the profile with boxcars of varying widths, this method does not depend on the shape of the profile. It is therefore applicable to more varying pulse morphologies and more in line with the S/N determination of algorithms, such as \heim, that search for SPs through convolving frequency-averaged data with boxcars. When the above injection-test procedure is repeated with a {\tt Furby} equivalent S/N scaling method, the recovered S/N values are indeed more in line with their requested values\footnote{No one-to-one agreement is found, as the test was performed with integer data of finite time resolution that can only represent an SP to a certain degree. The recovered S/N agreement may therefore vary depending on the time resolution of the integer-type data used.} (Fig. \ref{fig:snr_recovery}, boxcar convolution scaling). Because of these two reasons, the S/N boxcar scaling method is implemented as the default method for the {\tt FRBfaker} after the injection tests were completed.

However, while the boxcar convolution scaling does not depend on the morphology of an injection, it does depend on the assumed search template of the algorithm used to search for SPs. The boxcar is the most commonly used search template among the SP search algorithms, therefore this method will give accurate reconstructed S/N values in most cases. To provide users with alternatives that might better retain connection to physical properties, two more methods were implemented to calculate a pulse profile's area other than through boxcar convolutions. These methods determine the profile's width and amplitude from which its area is deduced in the Furby equivalent S/N scaling method. The first (indicated as the `box\_est\_wth' method in Fig. \ref{fig:snr_recovery}) determines the profile's width by assessing how many of its bins have a value above zero and then calculates the profile's amplitude by dividing the sum of the profile by this width. The second method (the `box\_est\_amp' method) takes the maximum value of the profile as its amplitude and then divides its sum by this amplitude to obtain the profile's equivalent width. These methods are less tailored towards the way \heim\ searches for SPs and thus will show a larger offset in the recovered S/N values for the S/N scaling test performed, as can be seen in Fig. \ref{fig:snr_recovery}. Which scaling method can best be used depends on the use case of the injections.

\subsection{Zero-DM filtering and bandpass removal test}
\citet{pab+2018} report that, for large pulse widths and decreasing DM, the power of their injected SPs is reduced due to the applied zero-DM filter used to remove broadband RFI from their data. A zero-DM filter is also applied to the HTRU-North data. To see if the brightness of injected SPs in the HTRU-North data have been reduced due to this filter, this quick test is performed. Two copies of the dataset were made that contain injections scaled with the `original {\tt FRBfaker} S/N scaling' method created for the S/N scaling test in Appendix \ref{sec:A.scaling_test}. Using {\tt RFIbye}, a zero-DM filter is applied on the first copy and the (flat) bandpass is removed from the second. Then, \heim\ is run over the data to determine the injections' resultant S/Ns. Indeed, the operations do reduce the recovered S/Ns of the injected SPs slightly (< 0.2\%). However, for real data with non-Gaussian statistics and wider bursts, these operations might have a larger effect. In this study, these losses are, however, assumed to be small and could be ignored.
\vfill

\subsection{Re-creation of known SPs test}\label{sec:A.re-creation}
The last test determines the degree to which existing SPs can be re-created with the {\tt FRBfaker}. That is, its ability is tested to create synthetic bursts with complex morphologies from a set of observational parameters. Therefore, three published (complex) bursts have been selected, from three different observatories, to be replicated and injected into synthetic data resembling the base data of the original burst. These bursts are bursts VII, C232, and FRB\,20191221A, observed with Effelsberg, Arecibo, and CHIME and published in \citet{hst+2019,jsn+2022,tab+2022}, respectively. The {\tt FRBfaker} versions of these bursts can be seen on the left-hand side of figures \ref{fig:A.eff}, \ref{fig:A.arecibo} and \ref{fig:A.chime}. While for comparison, the original bursts are shown on the right of these figures in similar colour maps and scales, to more easily identify their resemblance. Since the S/N values reported for the published bursts have been determined with boxcar convolving search algorithms, the synthetically created bursts are injected using the `boxcar convolution scaling' method to scale their power. The similarity of the synthetic and real bursts across the different observatories shows the general applicability of the {\tt FRBfaker} that makes it suitable for a large range of use cases.

\end{multicols}

\subsubsection{Effelsberg}\label{sec:A.effelsberg}
\begin{figure*}[h!]
    \centering
    \includegraphics[width=0.9\textwidth]{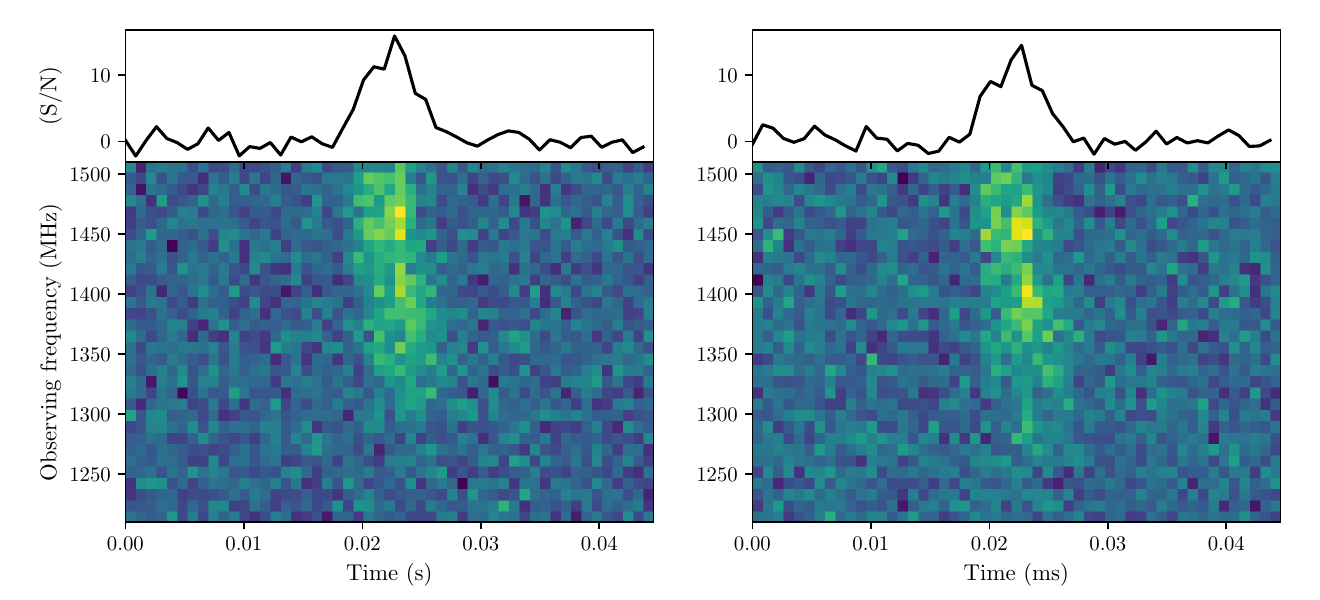}
    \caption{Dynamic spectra from the re-creation test of burst VII \citep{hst+2019} observed with Effelsberg. Left: {\tt FRBfaker} replicate of the burst. Right: Original burst's dynamic spectrum using the same colour map and scaling as the replicated burst.}
    \label{fig:A.eff}
\end{figure*}

Published in \citet{hst+2019}, burst VII of FRB\,20121102A was selected due to its relatively low S/N while still showing structure in its dynamic spectrum. By least-square fitting 2D Gaussians to the data, four burst subcomponents were identified from the burst's original dynamic spectrum. These four subcomponents are injected in integer value filterbank data, comprised of pure Gaussian noise, with the same data resolution as that of the HTRU-North data and scaled to an integrated S/N of 28 as published in \citet{hst+2019}. To attain the synthetic pulse profile as shown on the left-hand side of Fig. \ref{fig:A.eff}, the injection was repeated several times. Due to its relatively low S/N and injection into integer-type data, the shape of its pulse profile is strongly affected by the base data values. Several injections were therefore needed to obtain a close match between the synthetic and original pulse profile of burst VII.

\begin{verbatim}
FRBfaker.py
    --Tprof "0.20506,6.99135,0.00568,0.20187,7.46921,0.00195,0.20443,9.52629,0.00236,
             0.20495,3.95085,0.00103"
    --Fprof "1369.375,112.5,1473.18083,96.31098,1470.31404,78.48546,1402.54621,29.76976"
    -d 560 -t 1 --snr 28 --plot_prof -v -o Effelsberg_remake_VII.fil fake --Ttot 5
\end{verbatim}

\newpage
\subsubsection{Arecibo}\label{sec:A.arecibo}
\begin{figure*}[h!]
    \centering
    \includegraphics[width=0.9\textwidth]{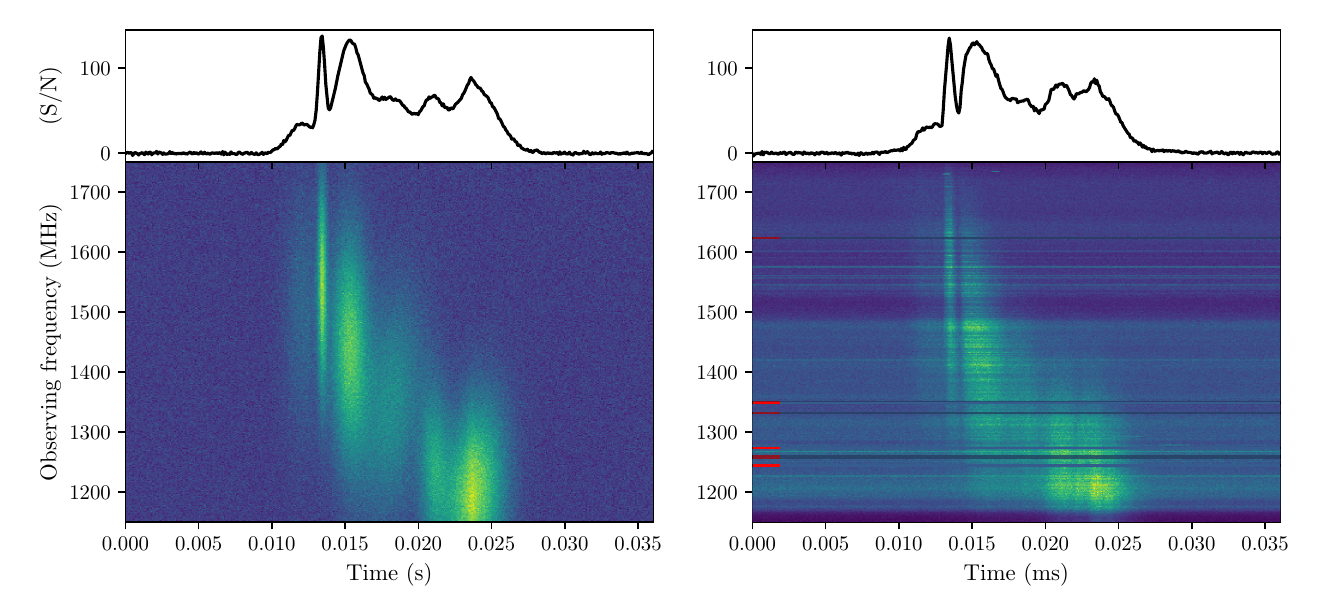}
    \caption{Non-normalised dynamic spectrum of burst C232 \citep{jsn+2022} on the right and its synthetic counterpart on the left. Differences in the dynamic spectra of the bursts are mainly due to the effects of the receiver bandpass still being present in the original data, while the replicated burst is injected in pure Gaussian noise.}
    \label{fig:A.arecibo}
\end{figure*}

For Arecibo, a very high S/N burst was selected, observed during the November burst storm of FRB\,20121102A in 2018, to see to what extent (precision) the {\tt FRBfaker} can replicate bursts. The parameters of the seven subcomponents of burst C232 were taken from Table C1 in \citet{jsn+2022} and converted to values that can be passed to the {\tt FRBfaker}. While the parameter values in Table C1 are obtained by least-square fitting 2D Gaussians to the non-normalised Arecibo data, the raw, non-normalised dynamic spectrum of the original burst is shown in Fig. \ref{fig:A.arecibo}. The frequency-averaged TSs of both the synthetic and original bursts are almost identical. The small discrepancies seen are due to bandpass effects that are not present in the synthetic burst, and one original subcomponent deviating slightly from having a pure Gaussian-shaped frequency-time profile. The majority of the complexity of burst C232 has otherwise perfectly been replicated with the {\tt FRBfaker}.

\begin{verbatim}
FRBfaker.py
    --Tprof "0.28285,12.00158,0.0021582,0.28415,37.97083,0.00050785,0.28598,40.79733,0.0020817,
             0.28883,21.89670,0.0037702,0.29174,14.88859,0.0015218,0.29437,3.68356,0.00046769,
             0.29445,26.66038,0.0035194"
    --Fprof "1503.63101,349.17298,1551.58302,285.20890,1448.99894,308.68704,
             1344.88792,303.46983,1244.95968,216.21944,1028.09978,479.69481,
             1198.27838,245.42009"
    --amp2snr -t 1 --tilt [3.12918,0.75455,3.17946,9.16823,0.95471,0.99126,6.23226]
    --plot_prof -v -o Arecibo_remake_C232.fil
    fake --dt 0.00001024 --fch1 1750 --chanbw 1.5625 --nchan 385 --MED 26 --RMS 15 --Ttot 3
\end{verbatim}

\newpage
\subsubsection{CHIME}\label{sec:A.chime}
\begin{figure*}[h!]
    \centering
    \includegraphics[width=0.9\textwidth]{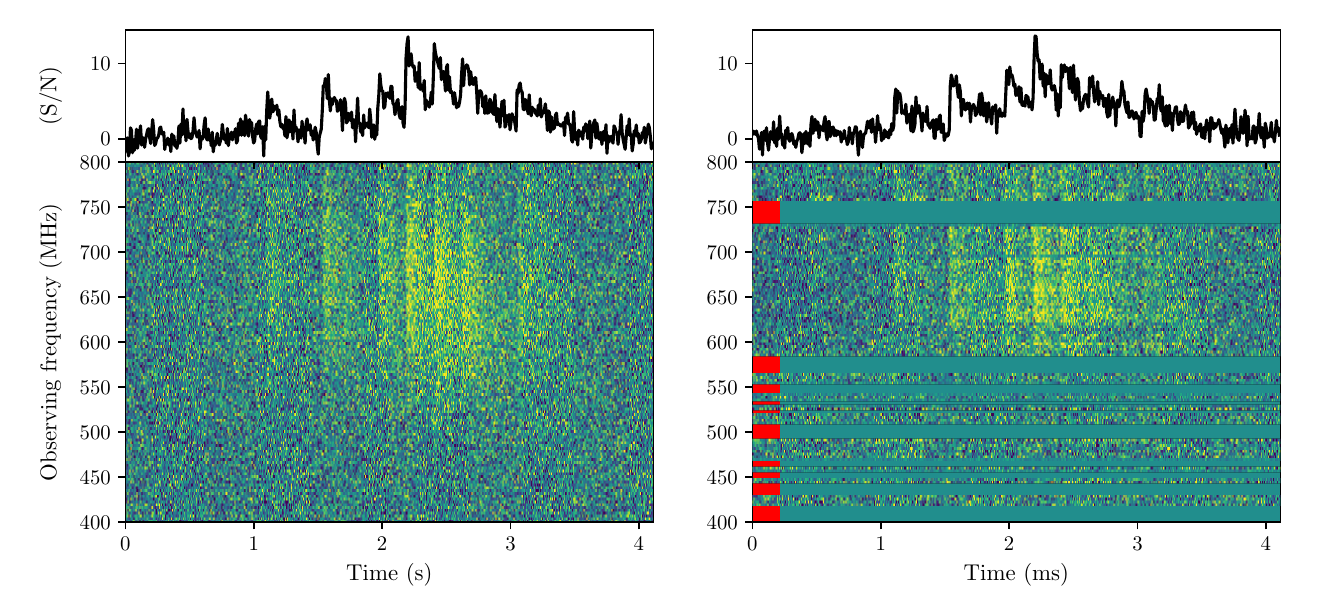}
    \caption{Dynamic spectra from the re-creation test of FRB\,20191221A. Left: Synthetic version of the burst created with the {\tt FRBfaker} and injected in synthetic data resembling that of CHIME. Right: Real burst that the synthetic version mimics. For both plots the same colour map and scale are used as in the burst's original publication \citep{tab+2022}.}
    \label{fig:A.chime}
\end{figure*}

Lastly, a burst detected with CHIME was chosen for this test because of CHIME's importance to the FRB field, as it has detected most of the known FRBs. Of all its detections, FRB\,20191221A did stand out as being the first observed FRB for which a period was found among its subcomponents. While this burst was mentioned to originate from J0248+6021, it is used here as an example of how well the complexity of SPs can be reconstructed with {\tt FRBfaker}, using their published subcomponent properties. It is therefore irrelevant whether the SP originated from an FRB or a pulsar. At the time of writing, no update about this SP has yet been published. Therefore, the reference to it as FRB\,20191221A is kept. To determine its period, this burst has been well modelled with models similar to those used by {\tt FRBfaker}. It can therefore easily be replicated using the model parameters published in \citet{tab+2022}, making it a good showcase. The only parameters not directly given are the S/N values of the burst's subcomponents, which have therefore been determined from the fit in figure 1, panel b, in \citet{tab+2022}. Both synthetic and real bursts agree well with one another, as can be seen in Fig. \ref{fig:A.chime}.

\begin{verbatim}
FRBfaker.py
    --Tprof "0,0.5388,0.004,0.43,0.5256,0.004,0.652,1.6870,0.004,
             1.0862,2.6897,0.004,1.52,2.95,0.004,1.736,3.8129,0.004,
             1.952,3.3487,0.004,2.171,2.7571,0.004,2.604,2.0487,0.004"
    --Fprof "650,200" -t 5 --scatter 0.34 600 --scat_dm 368 --amp2snr --abssnr 
    --plot_prof -v -o CHIME_remake_FRB20191221A.fil
    fake --dt 0.000983 --fch1 800 --chanbw 3.125 --nchan 128 --Ttot 12
\end{verbatim}

\newpage

\section{Injections in integer type data} \label{sec:A.int_inject}
\begin{figure*}[h!]
    \centering
    \includegraphics[width=0.78\textwidth]{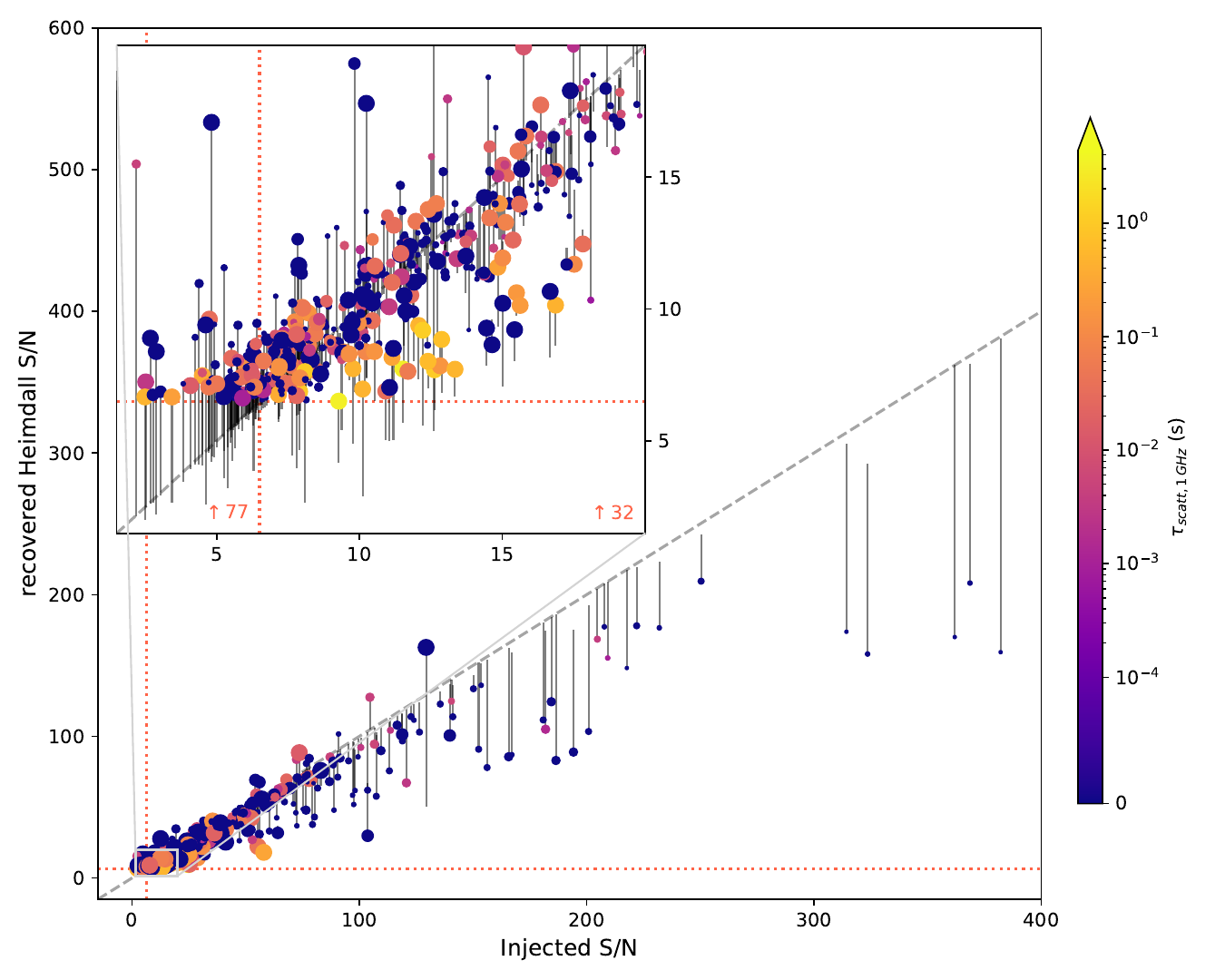}
    \caption{All detected injections with their injected S/Ns plotted against their recovered S/Ns by \heim. Marker sizes indicate \heim's reported width of the injections and the colours show the scattering times with which they are injected referenced to 1\,GHz. The vertical lines display a rise or drop in S/N between an SP's discrete injected S/N and the recovered \heim\ S/N. 109 injections (red numbers in inset) with a discrete injected S/N below the set S/N threshold of 6.5 (dashed red lines) were unexpectedly detected due to one of the reasons listed in Sect. \ref{sec:D-discrepancies}.}
    \label{fig:detected_injections}
\end{figure*}

\vspace{-1.1em}
\begin{multicols}{2}\noindent
Injection of SPs in integer type data influences their recoverable S/N (circumstance four in Sect. \ref{sec:D-discrepancies}). Due to the limited bit range of integer data, the precision with which injections can be represented is restricted. This is evident in Fig. \ref{fig:detected_injections} as brighter injections are recovered by \heim\ with an increasingly lower S/N than the injections' discrete injected S/N. When the {\tt FRBfaker} needs to inject a very bright pulse in integer data, it will scale its pulse profile to try to match the integer profile's area with that of its creation profile in floating point values. If due to this matching, data samples of its dynamic spectrum reach higher values than the maximum value of the data type's bit-range (255 for 8-bit data), the {\tt FRBfaker} will cap these samples to this maximum value and tries to raise the value of other samples to give the integer profile the correct area. However, the injection's dynamic spectrum consists of a limited amount of data samples and can therefore only attain a certain maximum area. A maximum S/N thus exists at which it can be injected in integer data that depends on the morphology of the injection. This explains why the injections with a large injected S/N and a relatively narrow width are recovered with much lower S/Ns than their discrete injected S/Ns in Fig. \ref{fig:detected_injections}, their profile's data samples reached the maximum value of the integer bit-range.

Likewise, the S/N values of very faint SPs can be boosted by injection into integer data. If, for instance, all data samples of a faint SP have decimal values below one in its creation profile, the {\tt FRBfaker} will start scaling the profile until enough of its data samples have attained an integer value above zero to give it roughly the same area as its creation profile. This process might change the morphology of an injection if its power is unevenly distributed across time and frequency. Only a limited number of samples, those with the highest initial values, might then be included in the integer profile. The integer converted profiles can be much narrower in (band)width than the creation profiles and can be recovered with higher S/N values.

The above has happened with all detections of strongly scattered injections with an injected S/N above eight, and that were not expected to be detected due to their discrete injected S/Ns lying below six and a half. Scattering has caused their power to be spread so thinly over their dynamic spectra's data samples that only the top parts of their profiles were injected. These higher frequency data samples retained the most power and were the first to attain a value above one due to the integer conversion of their profiles. Their dynamic spectra were so deformed with power concentrated in just a small part of the original bursts' morphology, boosting their recovered S/N. In the inset plot of Fig. \ref{fig:detected_injections}, the injections for which this occurred can be identified as those injections with a discrete injected S/N below six and a half, a large width, and scattering timescales above $\gtrsim$0.01\,s at 1\,GHz. Those injections with no scattering applied to them and which were unexpectedly detected are mainly detected due to circumstance six in Sect. \ref{sec:D-discrepancies}. The injection with the lowest discrete and injected S/N in the inset plot is detected due to circumstance two.
\end{multicols}

\vspace{-1em}

\twocolumn[\section{Additional figures and tables}]

\begin{table}[h!]
  \small
  \caption{All 141 SP candidates identified in the analysis PTs.}
  \label{tab:sp_candidates}
  \centering
  \begin{tabu}{l c c c r c c r}
    \hline
    \noalign{\smallskip}
    Name&GL&GB&S/N&DM&DM$_{Gal}$&Width&Mem.\\
    \rowfont{\tiny}
    & (deg) & (deg) && \multicolumn{2}{c}{(\dm)} & (ms) & \\
    \noalign{\smallskip}
    \hline
    \multicolumn{8}{c}{\ldots} \\
    \noalign{\medskip}
    C044 & 78.75 & 0.51 & 6.83 & 86.3 & 394.7 & 1.7 & 5 \\
    C045 & 78.88 & 0.43 & 7.00 & 36.8 & 396.5 & 0.9 & 7 \\
    C046 & 79.00 & 1.81 & 6.80 & 26.1 & 357.0 & 3.5 & 6 \\
    \multicolumn{8}{c}{\ldots} \\
    \noalign{\smallskip}
    \hline
  \end{tabu}
  \tablefoot{The full table is available at the CDS.}
\end{table}

\vfill

\begin{figure}[h!]
    \centering
    \includegraphics[height=0.32\textwidth,angle=-90]{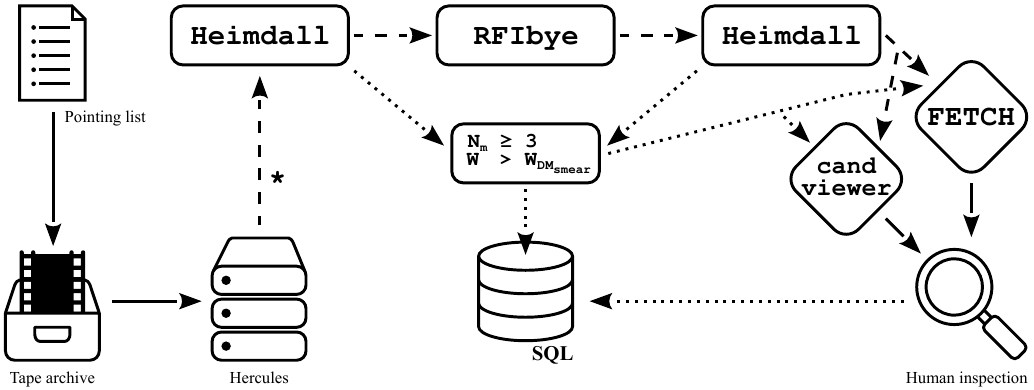}
    \caption{HTRU-North SP pipeline outlined in a flow diagram. Dashed lines represent the flow of filterbank files and the dotted lines indicate the processing of \heim\ candidate information. Intermediate and end results of the search are all stored in an SQLite database together with its progress. For the injection analysis, presented in Sect. \ref{sec:analysis}, synthetic SPs are injected at the starred processing step.}
    \label{fig:sp-pipeline}
\end{figure}

\begin{table}[ht]
\small
\caption{Pulse trains detected in the processed PTs.}
\label{tab:pulse_trains}
\centering
\begin{tabu}{l c c l l r r}
    \hline
    \noalign{\smallskip}
    & GLong & GLat & S/N & DM & Width & Mem.\\
    \rowfont{\tiny}
    & (deg) & (deg) & & (\dm) & \multicolumn{1}{c}{(ms)} & \\
    \noalign{\smallskip}
    \hline
    \noalign{\smallskip}
    \rowfont{\bfseries}
    \multirow{8}{*}{\rotatebox[origin=c]{90}{{\tiny T093.1-08.5}}} & 93.12 & -8.51 & & \multicolumn{3}{l}{37.8 $\pm$ 1.7 \quad P - 0.122\,(s)} \\
    \rowfont{\tiny}
    & \multicolumn{2}{r}{\textsc{i}.} & 6.80& 39.3 & 7.0 & 5 \\
    \rowfont{\tiny}
    & \multicolumn{2}{r}{\textsc{ii}.} & 6.96 & 38.3 & 0.4 & 3 \\
    \rowfont{\tiny}
    & \multicolumn{2}{r}{\textsc{iii}.} & 6.66 & 40.7 & 1.7 & $^{*}$2 \\
    \rowfont{\tiny}
    & \multicolumn{2}{r}{\textsc{iv}.} & 6.83 & 36.9 & 7.0 & 20 \\
    \rowfont{\tiny}
    & \multicolumn{2}{r}{\textsc{v}.} & 6.64 & 35.5 & 7.0 & 4 \\
    \rowfont{\tiny}
    & \multicolumn{2}{r}{\textsc{vi}.} & 7.12 & 36.9 & 7.0 & $^{*}$1 \\
    \rowfont{\tiny}
    & \multicolumn{2}{r}{\textsc{vii}.} & 6.64 & 37.1 & 7.0 & $^{*}$1 \\
    \noalign{\medskip}
    \hline
    \noalign{\smallskip}
    \rowfont{\bfseries}
    \multirow{5}{*}{\rotatebox[origin=c]{90}{{\tiny T102.1+01.4}}} & 102.12 & 1.44 & & \multicolumn{3}{l}{153.2 $\pm$ 9.8 \quad P - 0.019\,(s)} \\
    \rowfont{\tiny}
    & \multicolumn{2}{r}{\textsc{i}.} & 6.88 & 160.0 & 14.0 & 4 \\
    \rowfont{\tiny}
    & \multicolumn{2}{r}{\textsc{ii}.} & 6.96 & 156.4 & 14.0 & 5 \\
    \rowfont{\tiny}
    & \multicolumn{2}{r}{\textsc{iii}.} & 6.77 & 138.7 & 1.7 & 4 \\
    \rowfont{\tiny}
    & \multicolumn{2}{r}{\textsc{iv}.} & 6.91 & 157.6 & 14.0 & 10 \\
    \noalign{\medskip}
    \hline
    \noalign{\smallskip}
    \rowfont{\bfseries}
    \multirow{4}{*}{\rotatebox[origin=c]{90}{{\tiny T121.8-03.5}}} & \textsuperscript{\textdaggerdbl}121.75 & -3.53 & & \multicolumn{3}{l}{218.2 $\pm$ 11.8 \quad P - 0.387\,(s)} \\
    \rowfont{\tiny}
    & \multicolumn{2}{r}{\textsc{i}.} & 7.04 & 224.8 & 3.5 & 3 \\
    \rowfont{\tiny}
    & \multicolumn{2}{r}{\textsc{ii}.} & 6.95 & 204.6 & 1.7 & 3 \\
    \rowfont{\tiny}
    & \multicolumn{2}{r}{\textsc{iii}.} & 7.18 & 225.3 & 3.5 & 5 \\
    \noalign{\medskip}
    \hline
    \noalign{\smallskip}
    \rowfont{\bfseries}
    \multirow{5}{*}{\rotatebox[origin=c]{90}{{\tiny T065.1-10.7}}} & \textsuperscript{\textdaggerdbl}65.13 & -10.68 & & \multicolumn{3}{l}{177.4 $\pm$ 1.5 \quad P - 0.026\,(s)} \\
    \rowfont{\tiny}
    & \multicolumn{2}{r}{\textsc{i}.} & 6.74 & 178.1 & 14.0 & $^{*}$1 \\
    \rowfont{\tiny}
    & \multicolumn{2}{r}{\textsc{ii}.} & 6.88 & 175.5 & 7.0 & 3 \\
    \rowfont{\tiny}
    & \multicolumn{2}{r}{\textsc{iii}.} & 6.75 & 177.2 & 1.7 & 4 \\
    \rowfont{\tiny}
    & \multicolumn{2}{r}{\textsc{iv}.} & 6.85 & 179.0 & 7.0 & $^{*}$2 \\
    \noalign{\medskip}
    \hline
    \noalign{\smallskip}
    \rowfont{\bfseries}
    \multirow{6}{*}{\rotatebox[origin=c]{90}{{\tiny T052.3+07.7}}} & 52.25 & 7.72 & & \multicolumn{3}{l}{88.5 $\pm$ 8.9 \quad P - 0.237\,(s)} \\
    \rowfont{\tiny}
    & \multicolumn{2}{r}{\textsc{i}.} & 6.58 & 86.7 & 1.7 & $^{*}$1 \\
    \rowfont{\tiny}
    & \multicolumn{2}{r}{\textsc{ii}.} & 6.84 & 99.3 & 0.9 & 3 \\
    \rowfont{\tiny}
    & \multicolumn{2}{r}{\textsc{iii}.} & 7.01 & 94.8 & 0.2 & $^{*}$1 \\
    \rowfont{\tiny}
    & \multicolumn{2}{r}{\textsc{iv}.} & 6.57 & 85.4 & 1.7 & $^{*}$1 \\
    \rowfont{\tiny}
    & \multicolumn{2}{r}{\textsc{v}.} & 6.80 & 76.3 & 0.9 & 3 \\
    \noalign{\medskip}
    \hline
    \noalign{\smallskip}
    \rowfont{\bfseries}
    \multirow{5}{*}{\rotatebox[origin=c]{90}{{\tiny T059.5+00.8}}} & 59.5 & 0.8 & & \multicolumn{3}{l}{199.8 $\pm$ 23.2 \quad P - 2.858\,(s)} \\
    \rowfont{\tiny}
    & \multicolumn{2}{r}{\textsc{i}.} & 6.96 & 183.5 & 7.0 & 5 \\
    \rowfont{\tiny}
    & \multicolumn{2}{r}{\textsc{ii}.} & 6.88 & 209.2 & 14.0 & $^{*}$1 \\
    \rowfont{\tiny}
    & \multicolumn{2}{r}{\textsc{iii}.} & 6.72 & 178.5 & 3.5 & 3 \\
    \rowfont{\tiny}
    & \multicolumn{2}{r}{\textsc{iv}.} & 6.72 & 228.1 & 14.0 & $^{*}$1 \\
    \noalign{\medskip}
    \hline
    \noalign{\smallskip}
    \rowfont{\bfseries}
    \multirow{4}{*}{\rotatebox[origin=c]{90}{{\tiny T055.4+00.2}}} & \textsuperscript{\textdagger}55.38 & 0.15 & & \multicolumn{3}{l}{176.9 $\pm$ 17.3 \quad P - 0.247\,(s)} \\
    \rowfont{\tiny}
    & \multicolumn{2}{r}{\textsc{i}.} & 6.82 & 169.4 & 14.0 & 3 \\
    \rowfont{\tiny}
    & \multicolumn{2}{r}{\textsc{ii}.} & 7.09 & 164.4 & 14.0 & 3 \\
    \rowfont{\tiny}
    & \multicolumn{2}{r}{\textsc{iii}.} & 7.16 & 196.7 & 14.0 & 4 \\
    \noalign{\medskip}
    \hline
    \noalign{\smallskip}
    \rowfont{\bfseries}
    \multirow{4}{*}{\rotatebox[origin=c]{90}{{\tiny T054.9-03.0}}} & \textsuperscript{\textdagger}54.88 & -3.03 & & \multicolumn{3}{l}{199.7 $\pm$ 26.1 \quad P - 0.012\,(s)} \\
    \rowfont{\tiny}
    & \multicolumn{2}{r}{\textsc{i}.} & 7.13 & 186.2 & 14.0 & 35 \\
    \rowfont{\tiny}
    & \multicolumn{2}{r}{\textsc{ii}.} & 7.43 & 183.0 & 7.0 & 11 \\
    \rowfont{\tiny}
    & \multicolumn{2}{r}{\textsc{iii}.} & 7.70 & 229.8 & 3.5 & 27 \\
    \noalign{\medskip}
    \hline
\end{tabu}
\tablefoot{
\begin{itemize}
    \item[$^{\dagger}$] Might be a known pulsar.
    \item[$^{\ddagger}$] Probably slightly outside the Galaxy according to the NE2001 model, but within 1.5 times DM$_{gal}$ for the specified direction.
    \item[$^{*}$] Non-valid candidate, and although its chance of being real is low, it has a morphology similar to the valid SPs of the pulse train and is therefore taken into account.
\end{itemize}
}
\end{table}

\end{appendix}
\end{document}